\documentclass[aps,reprint,superscriptaddress,longbibliography,amsmath,amssymb]{revtex4-2}

\usepackage{graphicx}
\usepackage{bm}
\usepackage{xcolor}
\usepackage[colorlinks=true,linkcolor=blue,citecolor=blue,urlcolor=blue]{hyperref}

\graphicspath{{.}}
\newcommand{\angstrom}{\textup{\AA}}

\begin{document}

\title{Learning and retrieval for warm-starting charge-self-consistent DFT+DMFT}

\author{Rishi Rao}
\affiliation{Department of Physics, Rutgers University, Newark, New Jersey 07102, USA}
\author{Li Zhu}
\email{li.zhu@rutgers.edu}
\affiliation{Department of Physics, Rutgers University, Newark, New Jersey 07102, USA}

\date{\today}

\begin{abstract}
Charge-self-consistent (CSC) DFT+DMFT delivers quantitative correlated-electron physics one configuration at a time, making ensemble sampling dependent on reliable warm starts for its expensive fixed-point iteration. Here we compare retrieval from the most similar converged structure with an E(3)-equivariant model that predicts a physics-structured self-energy and Fermi level. Because the full CSC loop refines both initializations, every production result remains a converged DFT+DMFT solution. Across metallic Fe, correlated FeO, and Mott-insulating NiO, learning reduces the typical iterations to sustained convergence from 8 to 3, 8 to 3, and 6 to 1. Retrieval matches this median speed when a dense same-state archive is available, but several poor transplants reveal that structural similarity does not predict transplant quality. In a pre-registered volume window excluded from training and donor pools, learning retains its speed, whereas retrieval starts $3.4\times$ farther from the fixed point and approaches cold-start cost. Either initialization can select a distinct near-degenerate branch on a rugged CSC landscape. Applied end to end, the workflow generates over one thousand correlated energy and force labels for iron at Earth's-core conditions and trains an equivariant interatomic potential. Solid--liquid coexistence with 9216 atoms gives $T_m = 6225 \pm 42\,\mathrm{K}_{\mathrm{stat}}$ at 330~GPa, consistent with recent experiments. A 50-configuration DFT+DMFT audit resolves the potential's energy calibration but does not justify a corrected melting temperature, because four-atom cells cannot realize a liquid. The resulting regime map favors retrieval within dense coverage, amortized learning at and beyond its boundary, and solver refinement throughout.
\end{abstract}

\maketitle

\section{Introduction}
\label{sec:intro}

Dynamical mean-field theory (DMFT) coupled to density-functional theory (DFT) is the workhorse for realistic simulations of correlated electron systems~\cite{dmftKotliar2006}. It captures local dynamical correlations non-perturbatively, and it gives access to quasiparticle renormalization, finite-temperature magnetism, and spectral properties across the metal--insulator landscape~\cite{mitAbrahams1996,mitMasatoshi1998,tmoAnisimov1997,tmoTokura2000,dmftArpita2019}. Its cost profile, however, is dominated by the quantum impurity problem. Each charge-self-consistent (CSC) DFT+DMFT calculation must iterate expensive impurity solves, continuous-time quantum Monte Carlo (CTQMC) in the general finite-temperature case, until impurity and lattice quantities agree. A single converged configuration is routine. The thousands of configurations demanded by ensemble averages, free energies, and phase boundaries are not. This gap between single-point capability and thermodynamic sampling has kept correlated-electron corrections out of most finite-temperature materials physics.

Machine learning has begun to attack this bottleneck from several directions. Surrogate impurity solvers and learned corrections improve approximate Green's functions~\cite{mlgfArsenault2014,mlgfSturm2021,mlgfDong2024,mlgfMitra2025,valenti2026neural,nain2026neural}; data-driven compression has also reduced the variational space of quantum-embedding solvers~\cite{giuli2026linear}. These studies establish that key many-body quantities are learnable, but they mostly act downstream of a specified impurity or embedding Hamiltonian. Their applications are commonly model systems or zero-temperature settings, and the learned output replaces part of the many-body calculation instead of feeding a verified one. Ensemble production poses a different choice: reuse the closest converged self-energy in a growing archive, or learn a structure-to-self-energy functional. To our knowledge, these alternatives have not been compared directly. Because both terminate in the same controlled solver, the distinction is operational rather than one of final theory. We find that the answer depends on coverage and, in doing so, identifies where learning adds value to a many-body workflow.

\begin{figure*}[t]
  \centering
  \includegraphics[width=1.0\textwidth]{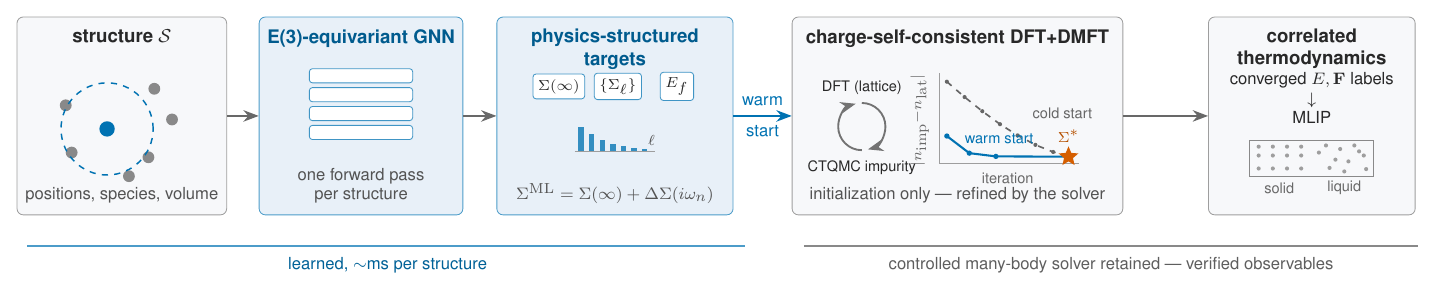}
  \caption{\textbf{Learned initialization of a verified many-body fixed-point solver.} An atomic configuration $\mathcal{S}$ (positions, species, volume) is mapped by E(3)-equivariant graph neural networks to a physics-structured set of targets: the static high-frequency limit $\Sigma(\infty)$, the Legendre coefficients $\{\Sigma_\ell\}$ of the dynamical part, and the Fermi level $E_f$, from which $\Sigma^{\mathrm{ML}}(i\omega_n)$ is reconstructed. The prediction initializes the charge-self-consistent DFT+DMFT loop, which retains its CTQMC impurity solver and convergence criteria; the controlled solver refines it to a converged solution in fewer iterations. On landscapes with a unique attractor this is the same fixed point $\Sigma^{*}$ that a cold start reaches. On multistable landscapes the initialization can select among near-degenerate branches (Sec.~\ref{sec:fixedpoint}). The converged energies and forces train an interatomic potential for large-scale correlated thermodynamics.}
  \label{fig:schematic}
\end{figure*}

We implement both strategies and compare them on the same structures. The learned initializer maps a crystal structure $\mathcal{S}$ to $\{\Sigma^{\mathrm{ML}}(\infty),\Sigma^{\mathrm{ML}}_\ell,E_f^{\mathrm{ML}}\}$ with E(3)-equivariant graph neural networks~\cite{e3nnGieger2022,e3nnKondor2018,e3nnThomas2018,e3nnWeiler2018}. Its targets form a compact, physics-structured representation of the orbital-resolved self-energy and Fermi level. Specifically, an exact static high-frequency limit is separated from a dynamical part expressed in a rapidly decaying Legendre basis~\cite{Boehnke2011,sigmaWang2011}. The model thus predicts a small set of real numbers per orbital rather than a noisy complex function on a dense Matsubara grid. Retrieval instead copies the converged $\{\Sigma,\mu\}$ from the fingerprint-nearest solved structure~\cite{fpZhu2016}, subject to pre-registered donor rules (Appendix~\ref{app:baselines}). Either output only initializes the full CSC cycle (Fig.~\ref{fig:schematic}); the original convergence criteria and CTQMC solver remain in control. Every production label is consequently a refined DFT+DMFT solution. Across Fe, FeO, and NiO, learning reduces typical iteration counts from 8 to 3, 8 to 3, and 6 to 1. Retrieval matches the median within a dense same-state archive but has a heavier tail, and its applicability ends with that coverage.

We then test what this acceleration enables at the application scale. Iron at Earth's-core conditions is demanding because its melting curve near 330~GPa sets the inner-core-boundary temperature. Experiments approach this pressure by extrapolation from either side~\cite{emcAnzellini2013,emcKraus2022,emcBalugani2024,emcSinmyo2019}, while DFT predictions span almost $10^3$~K~\cite{aimcAlfe2002,aimcAlfe2009,aimcBelonoshko2021,aimcBelonosko2000,aimcGonzalez2023,aimcSola2009,aimcStixrude2014,aimcSun2022,aimcSun2023,aimcWu2024,fepSun2018}. Dynamical correlations also differ between the solid and liquid phases~\cite{ieccPourovskii2017,ieccPourovskii2019,ieccVekilova2015}. The warm-started workflow expands the corpus to 1052 converged DFT+DMFT energy--force labels across solid and liquid Fe configurations, which train an equivariant MLIP. We use it in 9216-atom coexistence simulations to determine the hcp-Fe melting curve. Dedicated DFT+DMFT ensembles at the measured coexistence state then probe the potential's calibration at the free-energy level. The production protocol gives $T_m=6225\pm42\,\mathrm{K}_{\mathrm{stat}}$ at 330~GPa, consistent with the most recent experimental constraints. To our knowledge, this is the first melting curve at these conditions based on DFT+DMFT energies and forces.

\section{A structure-conditioned, physics-structured self-energy functional}
\label{sec:method}

Within DFT+DMFT, electronic correlations enter entirely through the local self-energy. For each correlated atom and orbital, the DMFT loop determines a frequency-dependent $\Sigma(i\omega_n)$ that, together with the lattice Green's function, fixes the electronic structure and thermodynamics. Converging it means finding the fixed point at which impurity and lattice quantities (occupancies, local Green's functions) agree. Formally, the converged self-energy is a functional of the local environment. It depends on the atomic configuration, composition, and thermodynamic state, as well as on the correlated subspace and interaction parameters. We make this viewpoint operational and seek an approximate, parameterized functional $\Sigma^{\mathrm{ML}}[\mathcal{S}](i\omega_n)$ mapping a structure $\mathcal{S}$ to an orbital-resolved self-energy close to the fixed point. The goal is not to replace the DMFT solution but to start the iteration near it.

\subsection{Representation}

\begin{figure*}[t]
  \centering
  \includegraphics[width=0.98\textwidth]{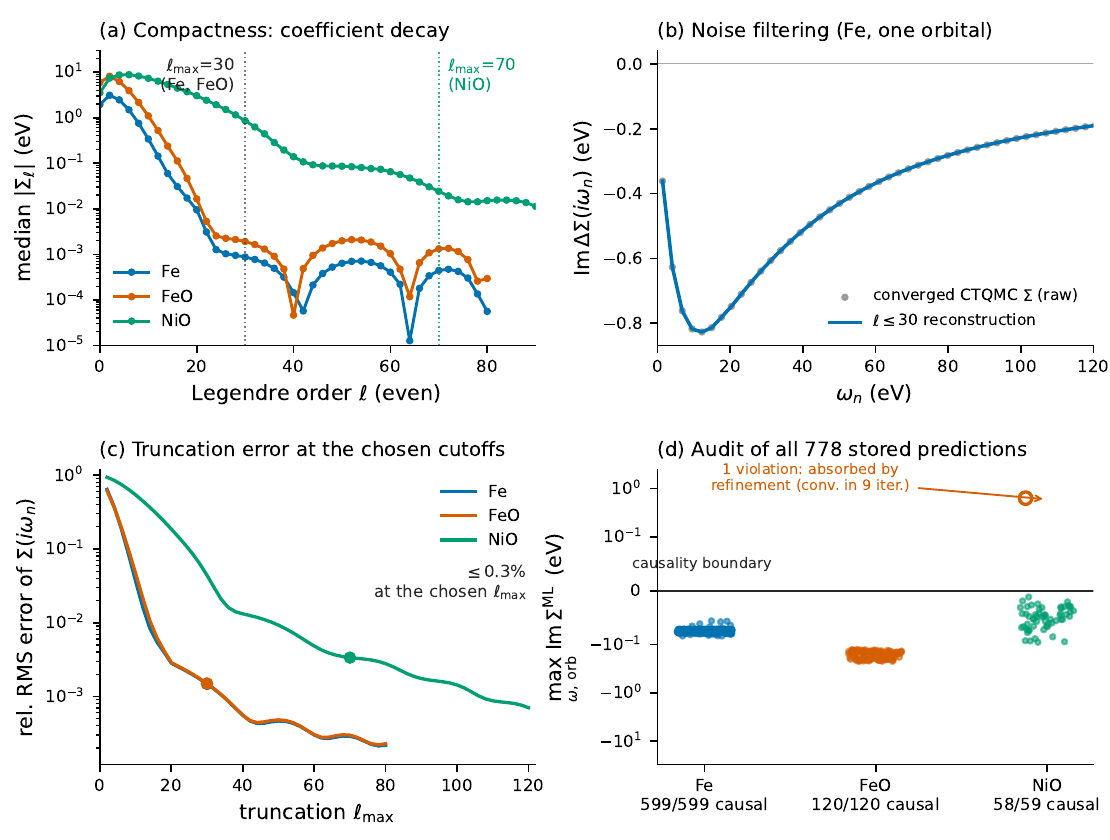}
  \caption{\textbf{The physics-structured representation is compact, noise-filtering, and audited.} (a) Median magnitude of the Legendre coefficients $|\Sigma_\ell|$ (even $\ell$) of the dynamical self-energy for converged solutions of Fe, FeO (both 5000~K), and NiO (611~K); dotted lines mark the truncations used. The slower decay of FeO and NiO reflects, respectively, stronger hybridization-induced frequency dependence and finer low-temperature structure. (b) The truncated representation filters CTQMC noise by construction: raw converged $\mathrm{Im}\,\Delta\Sigma(i\omega_n)$ for one Fe $d$ orbital (points) versus its $\ell \le 30$ reconstruction (line), which also recovers the exact high-frequency limit. (c) Relative RMS error of the reconstructed $\Sigma(i\omega_n)$ versus truncation: for the hot systems the error saturates at the CTQMC label-noise floor by $\ell \approx 20$; for NiO it continues to decrease through $\ell_{\max}=70$ (marker). (d) Causality audit of every stored production prediction of this work: the maximum of $\mathrm{Im}\,\Sigma^{\mathrm{ML}}$ over Matsubara frequencies and orbitals, per structure. 777 of 778 predictions are strictly causal; the single violation (one NiO structure, $0.04\%$ of grid points) was repaired by the downstream self-consistency, which converged in 9 iterations. Prediction error costs iterations rather than observables wherever the charge-self-consistent landscape has a unique attractor; branch selection on multistable landscapes is characterized in Sec.~\ref{sec:fixedpoint}.}
  \label{fig:representation}
\end{figure*}

A naive strategy, regressing $\Sigma(i\omega_n)$ directly on a Matsubara grid, is poorly conditioned. The CTQMC self-energy is complex-valued and noisy at high frequency. It also obeys strict analytic requirements, including a well-defined high-frequency expansion $\Sigma(i\omega_n) = \Sigma(\infty) + \mathcal{O}(1/i\omega_n)$~\cite{sigmaWang2011} and a definite sign of the imaginary part on the Matsubara axis. We instead adopt a representation structured by this physics. The static limit $\Sigma(\infty)$, which carries the Hartree--Fock-like contribution of the local interaction, is separated exactly,
\begin{equation}
\Sigma(i\omega_n) = \Sigma(\infty) + \Delta\Sigma(i\omega_n),
\end{equation}
and the dynamical part, which decays at high frequency, is Fourier transformed to imaginary time and expanded in a Legendre basis~\cite{Boehnke2011},
\begin{equation}
  \Delta\Sigma(\tau) = \sum_{\ell=0}^{\ell_{\max}} \sqrt{2\ell+1}\, P_\ell(x(\tau))\, \Sigma_\ell,
  \qquad x(\tau) = \tfrac{2\tau}{\beta} - 1 .
\end{equation}
The real coefficients $\Sigma_\ell$ decay rapidly with $\ell$ [Fig.~\ref{fig:representation}(a)]. At 5000~K, $\ell_{\max}=30$ suffices for Fe and FeO; NiO at 611~K requires $\ell_{\max}=70$ to resolve finer low-temperature structure (Supplemental Note~7). The targets are therefore a small set of real numbers per orbital and site. Reconstruction restores the exact high-frequency limit by construction and filters the Monte Carlo noise that a gridwise fit would retain [Fig.~\ref{fig:representation}(b)]. We fixed the truncations by convergence rather than model tuning. At these cutoffs, the stored self-energies are reproduced with relative RMS errors of $0.15\%$ for Fe and FeO and $0.34\%$ for NiO [Fig.~\ref{fig:representation}(c)]. Those values lie one to two orders of magnitude below the learned prediction errors, leaving the representation negligible in the initialization error budget.

The representation enforces the high-frequency limit and a compact real-valued form. A truncated Legendre expansion does not, however, enforce causality (sign-definiteness of $\mathrm{Im}\,\Sigma$ on the Matsubara axis) or all Hermiticity and symmetry relations. We therefore call the representation physics-structured rather than physics-constrained and validate the residual properties explicitly. Figure~\ref{fig:representation}(d) reports this audit across all 778 stored production predictions (599 Fe, 120 FeO, 59 NiO). Of these, 777 are strictly causal at every Matsubara frequency and orbital, with $\mathrm{Im}\,\Sigma^{\mathrm{ML}}$ bounded away from zero. The single exception was a NiO structure that violated causality on $0.04\%$ of its grid points by at most $+0.64$~eV. The loop refined this initialization without incident and reached a causal fixed point in 9 iterations. Because every warm start undergoes full self-consistency, imperfect causality can increase the iteration count but cannot contaminate a production observable.

DMFT calculations additionally require the Fermi level $E_f$ to enforce the total-charge constraint; errors in $E_f$ produce impurity--lattice occupation mismatches that slow or prevent convergence. We therefore treat $E_f$ on the same footing and learn the joint map $\mathcal{S} \mapsto \{\Sigma^{\mathrm{ML}}(\infty), \Sigma^{\mathrm{ML}}_\ell, E_f^{\mathrm{ML}}\}$, a complete set of initial conditions for the CSC loop.

\subsection{Equivariant architecture and the orbital frame}

The functional uses E(3)-equivariant message-passing networks implemented in \texttt{e3nn}~\cite{e3nnGieger2022}; Appendix~\ref{app:network} gives the architecture and training details. Orbital-specific heads return $\Sigma(\infty)$ and the Legendre coefficients for each inequivalent correlated atom, while a pooled scalar head returns $E_f$. The five $d$ components are labeled in the fixed global real-harmonic frame ($z^2$, $x^2{-}y^2$, $xz$, $yz$, $xy$) used by the eDMFT projectors. Predictions and targets are therefore consistent within a common frame, although the composite map to frame-labeled components is not exactly covariant under arbitrary rotations. Rotating a snapshot without its projector convention would mix the $d$ components in a way that the per-orbital scalar heads cannot follow. This limitation is immaterial for the present campaigns, whose training, test, and production configurations all share one global frame; transfer across disorder and phases remains within that convention.

We checked this internal consistency by rigidly rotating the cell and atomic positions together, which is equivalent to co-rotating the projector frame. For five structures and three random rotations per compound, $\Sigma_\ell$ changed by at most $2\times10^{-5}$~eV, at the level of single-precision numerics. The changes in the $E_f$ and $\Sigma(\infty)$ heads were smaller still, no more than $8\times10^{-6}$~eV. Both are four orders of magnitude below the physical error scale. Exact covariance could be imposed with tensorial $\ell{=}2$ readouts contracted against the projector frame, without changing the basic architecture.

\section{Accuracy, acceleration, and verified refinement}
\label{sec:results}

\begin{figure*}[t]
  \centering
  \includegraphics[width=1.0\textwidth]{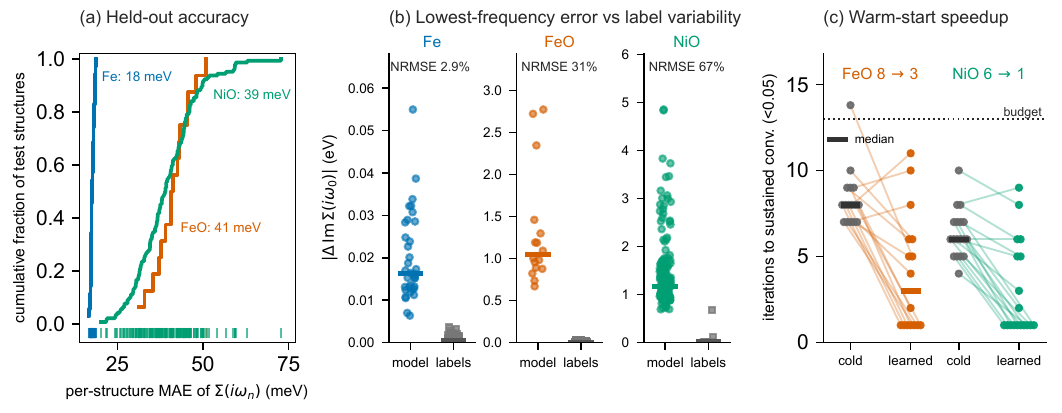}
  \caption{\textbf{Held-out accuracy and the low-frequency label-resolution audit.} (a)~Cumulative distributions of the per-structure MAE of the tail-subtracted $\Sigma(i\omega_n)$ on the exact published test splits (medians 0.018, 0.041, 0.039~eV for Fe $n=35$, FeO $n=16$, NiO $n=144$); ticks mark individual structures. (b)~Absolute prediction deviations at the lowest Matsubara frequency (active orbital components) beside the measured variability of the labels themselves: iteration-to-iteration differences of converged runs for Fe and FeO ($10^8$ Monte Carlo steps), and the deviation of the original $10^7$-step labels from dedicated $10^8$-step continuations for NiO. Relative accuracy over the first ten Matsubara points (band NRMSE, pre-declared window) is annotated. The labels are reproducible to $\lesssim 0.02$~eV in all three compounds, so the FeO and NiO low-frequency deviations are genuine model error, the hardest regime for the learned functional. (c)~Paired iterations to sustained convergence on hash-verified same-structure pairs: FeO $8\to3$ ($n=16$), NiO $6\to1$ ($n=18$); the compound with the largest low-frequency error retains the largest acceleration. Fe distributions use the retrained full-corpus control model of Sec.~\ref{sec:ood} (test accuracy matches the production model's published errors); FeO and NiO use the archived production weights. Fe's four-arm iteration comparison appears in Fig.~\ref{fig:verdict}; example spectra and historical averaged curves are in the Supplemental Material.}
  \label{fig:occconvergenceandsigpreds}
\end{figure*}

\begin{table}[b]
    \begin{center}
        \begin{tabular}{lcccccc}
         \hline\hline
          MAE (eV) & $E_f$ & $\Sigma_{z^2}$ &  $\Sigma_{x^2-y^2}$ & $\Sigma_{xz}$ & $\Sigma_{yz}$ & $\Sigma_{xy}$ \\
         \hline
         Fe & 0.217 & 0.302 & 0.298 & 0.315 & 0.310 & 0.319  \\
         FeO & 0.707 & 0.458 & 0.341 & 0.417 & 0.385 & 0.408 \\
         NiO & 0.208 & 0.198 & 0.200 & 0.378 & 0.387 & 0.314 \\
         \hline\hline
        \end{tabular}
        \caption{Mean absolute errors of predictions for $E_f$ and $\Sigma(\infty)$ orbital components on the held-out test sets. The operationally relevant question, how these errors propagate into iterations to convergence, is answered directly in Sec.~\ref{sec:accel}.}
        \label{tab:siginfandef}
    \end{center}
\end{table}

\subsection{Accuracy of the learned functional}

We test the functional against fully converged DFT+DMFT for metallic Fe and correlated B2 FeO at 5000~K, and for Mott-insulating NiO at 611~K. Rather than selected examples, Fig.~\ref{fig:occconvergenceandsigpreds} shows all 195 structures in the exact published test splits (Appendix~\ref{app:training}). The median Matsubara-grid errors are 0.018~eV for Fe, 0.041~eV for FeO, and 0.039~eV for NiO. A grid-wide percentage would be misleading because the tail-subtracted self-energy approaches zero at high frequency and makes the result denominator-dominated. We instead use a pre-declared band of ten low-frequency Matsubara points, where the relative errors are 2.9\%, 31\%, and 67\%, respectively.

An audit of the labels [Fig.~\ref{fig:occconvergenceandsigpreds}(b)] shows that these low-frequency deviations are model error rather than label noise. Converged-run iteration differences are below 0.02~eV for Fe and FeO. For NiO, five dedicated continuations at tenfold Monte Carlo statistics give a median label reproducibility of 0.014~eV, compared with a median model deviation of 1.1~eV. This result refutes our initial hypothesis that the NiO deviation arose from the $10^7$-step label statistics. Three continuations reproduce their labels to better than 0.015~eV, one differs by 0.12~eV, and one remains unstable at low frequency even at $10^8$ steps. The last is a hard case in the sense of Sec.~\ref{sec:fixedpoint}, rather than a statistical-noise artifact.

Low-frequency self-energies in the strongly hybridized and gapped compounds are therefore the hardest targets for the frame-fixed scalar readouts. By contrast, the exact symmetry zeros are reproduced without error. Full refinement absorbs this predictive error, and the same compounds retain substantial warm-start acceleration [Fig.~\ref{fig:occconvergenceandsigpreds}(c)]. Table~\ref{tab:siginfandef} reports the complementary orbital-resolved mean absolute errors for $\Sigma(\infty)$ and the error in $E_f$. The predictions provide effective starting guesses in all three compounds, and further data generation systematically improves them.

The distributions also expose two less favorable facts. For Fe, the Fermi-level error has a 0.17~eV median but a 0.70~eV 90th percentile. Thus the quantity most fragile under donor transplantation is also the least uniformly learned, although the CSC loop converts either error into extra iterations. In addition, the original Fe production weights were not archived, so the Fe panel uses the retrained full-corpus control model of Sec.~\ref{sec:ood}. Its median Legendre-coefficient error, 0.012 rather than the published 0.011, reproduces the production accuracy and closes the gap through a reproducible training pipeline.

\subsection{Learning versus retrieval: the four-arm comparison}
\label{sec:accel}

\begin{figure*}[t]
  \centering
  \includegraphics[width=0.98\textwidth]{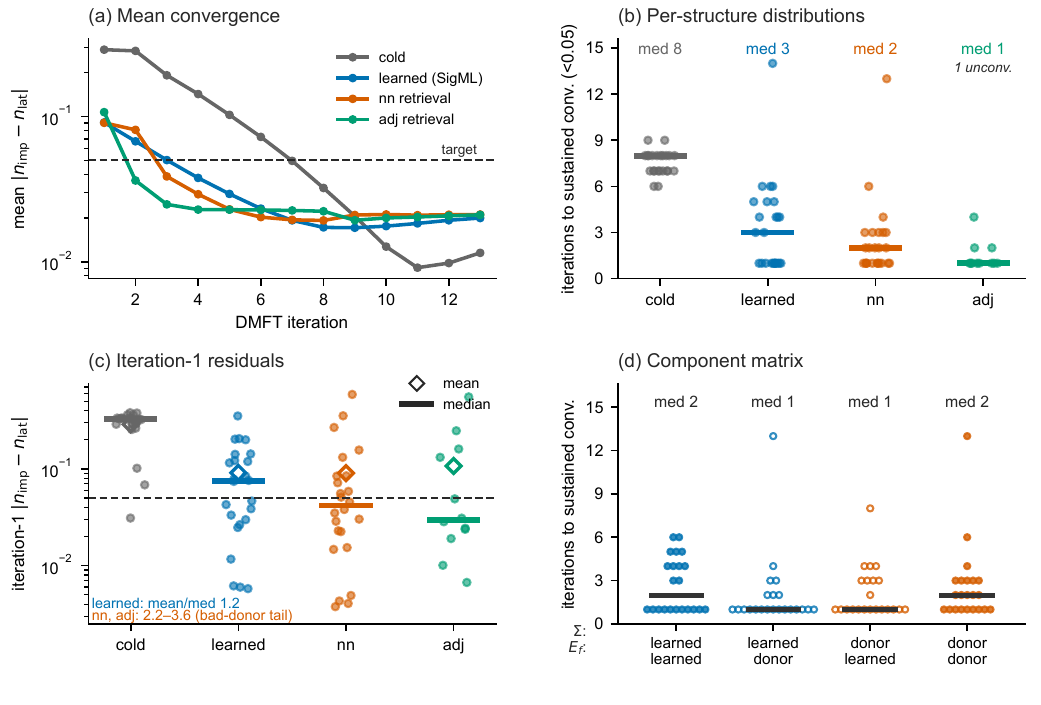}
  \caption{\textbf{Four initializations, same 24 structures, one criterion.} (a) Mean $|n_{\mathrm{imp}}-n_{\mathrm{lat}}|$ versus CSC iteration for cold ($\Sigma=0$), learned (SigML), fingerprint-nearest retrieval (nn), and nearest-sibling retrieval (adj) initializations of identical pre-registered Fe structures (12 distorted + 12 near-ideal). (b) Per-structure iterations to sustained convergence ($|n_{\mathrm{imp}}-n_{\mathrm{lat}}| < 0.05$ maintained to the end of the run); bars mark medians (8 / 3 / 2 / 1). (c) The residual each initialization lands at on iteration 1 (log scale; bar = median, open diamond = mean). Retrieval matches the learned model at the median but carries several poor transplants (mean/median 2.2--3.6 versus 1.2): donor similarity does not predict transplant quality, whereas no comparably poor transplant residual was observed in the learned arm on this set. (d)~The $2\times2$ initialization-component matrix (Sec.~\ref{sec:accel}) on the 22 donor-valid archive structures with complete quadruples: color marks the $\Sigma$ source, filled circles the pure arms, open circles the mixed arms ($\Sigma$ from one source, $E_f$ from the other); bars mark medians. Either mixture is as fast as either pure arm. This structure set differs from the 24-structure set of (a)--(c), so medians differ slightly.}
  \label{fig:verdict}
\end{figure*}

The operational figure of merit is the number of CTQMC-bearing iterations to convergence. A single crossing of the criterion can be transient; we observe cold runs that dip below 0.05 at iteration 1 before rebounding to 0.37. All counts below therefore use a sustained criterion: $|n_{\mathrm{imp}} - n_{\mathrm{lat}}| < 0.05$ maintained to the end of the run. The per-iteration fluctuation floor of this quantity at our CTQMC statistics is ${\approx}0.02$, so 0.05 is the tightest criterion that convergence can be meaningfully tested against per iteration. Production label runs additionally applied the charge-averaged stopping rule of Appendix~\ref{app:dmft}.

Figure~\ref{fig:verdict} compares four initializations on the same 24 pre-registered Fe structures, all run through an identical pipeline (Appendix~\ref{app:baselines}). Cold starts require a median of 8 iterations, with a range of 6--9 and no failures. Learning lowers the median to 3, although the distorted and near-ideal subsets differ sharply at 4.5 and 1, respectively. Fingerprint-nearest retrieval requires 2 iterations and nearest-sibling retrieval only 1. Within a dense, same-state archive, a nearby converged self-energy is therefore at least as effective as the learned prediction. This is a substantive property of the CSC problem: the fixed point is locally redundant in structure space, making a good donor nearly optimal.

The strategies separate in their tails rather than their medians [Fig.~\ref{fig:verdict}(c)]. Learning gives a compact iteration-1 residual distribution, with mean/median $=1.2$ and a median of 0.075. Both retrieval arms are right-skewed, with mean/median ratios of 2.2 and 3.6, because fingerprint distance does not determine transplant quality. One donor at distance 0.027, as close as several successful donors, produces an initialization worse than cold and then relaxes at the cold-start rate. The fragile quantity in this case is the transplanted chemical potential. Retrieval also needs an online archive of converged same-state solutions and a similarity search, resources absent at the start of a new campaign. Learning has its own prerequisite: our model began with roughly 500 cold-start labels. Within covered territory its advantage is therefore reliability, not speed; Sec.~\ref{sec:ood} tests what happens beyond that territory.

For FeO and NiO, structure hashes and Matsubara-grid checks identify genuinely paired cold and warm runs in the archive. Learning reduces the median from 8 to 3 iterations across 16 FeO pairs, and from 6 to 1 across 18 NiO pairs. Historical family averages show the same trend (Supplemental Material), but they mix structures across batches and overweight slow stragglers. Thus the typical FeO structure converges at iteration 3, whereas the averaged curve crosses only at iteration 5. The iteration-1 medians agree remarkably well across the pre-registered pairs, production history, and coverage-gap window: 0.055, 0.054, and 0.054. FeO remains the slowest learned start, consistent with the stronger frequency dependence induced by Fe~$3d$--O~$2p$ hybridization (Supplemental Note~7).

Cluster accounting places the four-month production label campaign at approximately $5.0\times10^{4}$ core-hours across all stages. Because those runs used fixed 20-iteration budgets, the acceleration appeared as convergence margin rather than a wall-clock reduction. Criterion-based stopping would reduce CTQMC solves by about 60\% per structure, from 8 to 3. The 13-iteration validation campaign, totaling $1.0\times10^{5}$ core-hours across all arms and follow-ups, partially realizes that saving. Retrieval and learning otherwise amortize nearly symmetrically: both consume the same converged corpus, training adds about one GPU-hour, and either query cost is negligible. Coverage and reliability, not cost, distinguish the two strategies.

To identify which component carries the acceleration, we crossed learned and retrieved values of $\Sigma$ and $E_f$ on 23 donor-compatible archive structures. One additional structure was excluded because its only donor used an incompatible single-impurity setup. The two mixed starts combine donor $\Sigma$ with predicted $\mu_0$, and predicted $\Sigma$ with donor $E_f$, using the same pipeline and criterion. The resulting matrix is nearly flat: either pure arm has a median of 2 iterations, whereas either mixture has a median of 1. Across the 22 complete quadruples, each mixture is faster than or tied with each pure arm in 16--21 cases. Any learned or retrieved self-energy therefore works with either Fermi-level estimate on this covered subset; neither component alone is fragile. The exception is the tail structure whose learned start requires 14 iterations. Its donor-$\Sigma$/predicted-$\mu_0$ mixture also misses sustained convergence within 13 iterations. The component test thus attributes the warm-start advantage to a jointly adequate $(\Sigma,E_f)$ pair, not to either entry by itself.

\subsection{Refinement, stationary branches, and multistability}
\label{sec:fixedpoint}

The initializer shortens the route to self-consistency but never replaces the controlled solver. The CSC loop retains its convergence criteria and CTQMC impurity solver, so a warm-started calculation remains a DFT+DMFT calculation in the strict sense. Every energy and force used downstream is therefore a refined DFT+DMFT result. An initializer need only be accurate enough to accelerate convergence; a poor initialization costs iterations rather than the correctness of the refined observable.

Refinement does not guarantee a unique answer on a rugged CSC landscape, although the extent of the problem depends strongly on the material. NiO is the simple case: all 17 verified warm--cold pairs agree at the solver-noise level. Their median $|\Delta E|$ is 1.3~meV/cell, with a 4.9~meV/cell 90th percentile, while $|\Delta\mu|$ has a 2.5~meV median. Impurity occupations agree to $6\times10^{-4}$.

Most FeO pairs behave similarly, with 13 of 16 giving a median $|\Delta E|\approx25$~meV/cell, but the remaining three settle on distinct solutions. Between branches, chemical potentials differ by 0.2--0.25~eV, per-Fe occupations by 0.03--0.05, and energies by 0.3--2.0~eV/cell. Both sides nevertheless satisfy the charge criterion at near-stationary energies. Fe shows the same pattern: 19 of 23 pairs agree, with median $|\Delta E|=54$~meV/cell, or 13.5~meV/atom. In the other four rattled structures, cold starts settle 0.5--2.2~eV/cell below the warm-started branches. Extending them to 21 iterations reduces the drift to $+0.3$--$0.9$~meV/atom per iteration without closing the separation. The difference is therefore not a settling transient.

Nor does either initialization consistently find the lower branch; in the largest FeO case, that branch comes from the warm start. We therefore refer to a solver-refined stationary branch rather than "the" fixed point. For strongly hybridized systems, an energy-stationarity check should accompany the charge criterion. The paired runs make visible a multistability that belongs to the charge-self-consistency map itself.

\subsection{Beyond coverage: a pre-registered coverage-gap interpolation test}
\label{sec:ood}

\begin{figure*}[t]
  \centering
  \includegraphics[width=0.98\textwidth]{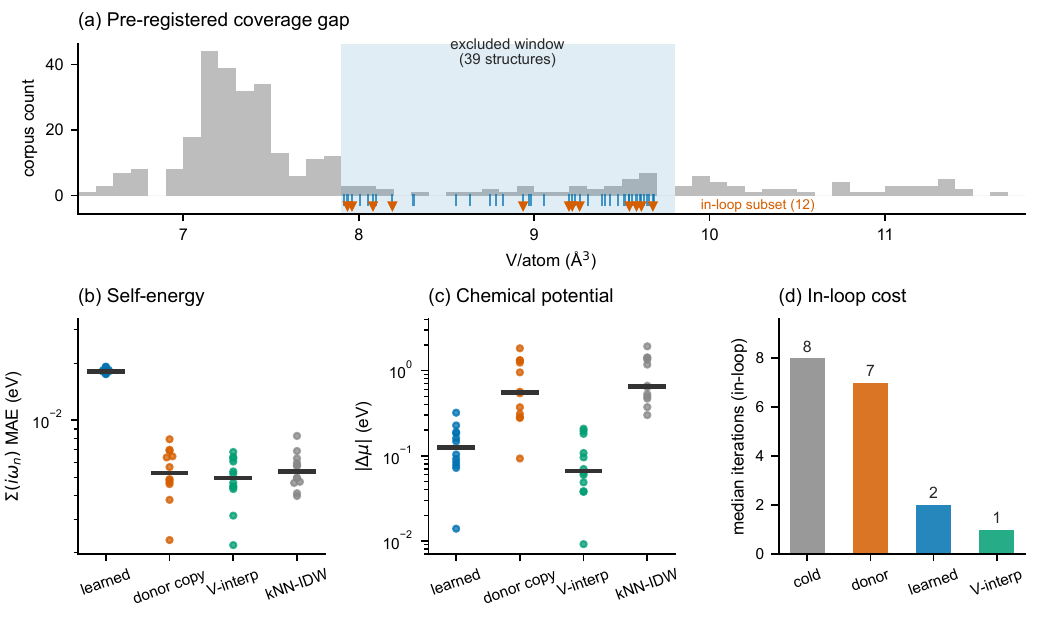}
  \caption{\textbf{The pre-registered coverage-gap interpolation test.} (a)~Volume distribution of the 350-structure training-era corpus. The shaded window ($V/\mathrm{atom} \in [7.9, 9.8]$~\AA$^3$, 39 structures, ticks) was excluded from the training set of the holdout model and from the donor pool; triangles mark the twelve structures given full in-loop DFT+DMFT tests. (b),(c)~Initializer quality on the twelve in-loop structures against the converged labels: the learned (holdout-excluded) model, the nearest eligible donor copied, linear interpolation in volume between the nearest eligible donors below and above the window, and inverse-distance-weighted $k{=}4$ averaging. Bars mark medians. Copying and interpolation beat the learned model on the self-energy (the converged $\Sigma$ is locally redundant); the copied chemical potential is the fragile element (median 0.55~eV), and volume interpolation repairs it (0.066~eV, below the learned model's 0.125~eV). (d)~Median in-loop iterations to sustained convergence for cold starts, nearest-donor transplants, the learned model (Sec.~\ref{sec:ood}), and the volume-interpolation arm (median 1; eleven of twelve sustained).}
  \label{fig:covgap}
\end{figure*}

Retrieval can choose only among discrete archived donors; whether learning bridges the space between them must be tested rather than assumed. Before training, we therefore pre-registered the coverage gap shown in Fig.~\ref{fig:covgap}(a). It contains 39 structures with $V/\mathrm{atom}\in[7.9,9.8]$~\AA$^3$, a sparse shoulder between the core-conditions region and the expanded-volume tail. We removed this window from both the retrained model's labels and the retrieval donor pool, recording the exclusion in a tamper-checked manifest (Appendix~\ref{app:ood}). An identically trained full-corpus model controls for retraining quality.

The model-side result is that the coverage gap costs little. On the 39 unseen-window structures, the holdout-excluded model's Legendre-coefficient error is 0.020 (median MAE), against 0.017 on its own in-distribution test. That is a degradation of only $1.2\times$ for entering a volume region it never saw. Against the 0.011 of the control model that trained on the window, the ratio is $1.8\times$, the price of coverage itself. Fermi-level errors remain at the production scale throughout (0.10--0.19~eV). The control model simultaneously certifies the retraining pipeline: its in-distribution accuracy matches the production model's published errors, so these comparisons measure coverage, not retraining quality. Retrieval cannot draw a donor from inside the window; its nearest eligible donors lie outside it, at fingerprint distances of 0.03--0.21 versus $\lesssim 0.03$ for typical in-distribution matches. The in-loop question is therefore what reaching across the gap costs.

We tested the in-loop cost with full CSC DFT+DMFT on a pre-registered random subset of twelve window structures, fixing the seed before any run. Each target used either the holdout-excluded model or its nearest eligible donor, under the production 13-iteration budget and sustained criterion of Sec.~\ref{sec:accel}. Learning converges eleven structures in a median of \textbf{2 iterations}, comparable to its in-distribution median of 3. Its iteration-1 residuals range from 0.03 to 0.17, with the same 0.075 median observed in distribution. Donor transplants also converge eleven structures, but require a median of \textbf{7 iterations}. Their initial residual has a 0.25 median, $3.4\times$ larger than learning, and spans 0.04--0.49. A node failure destroyed one donor run at iteration 3; the identical repeat converged cleanly at iteration 8. Retrieval therefore remains viable but surrenders most of its acceleration, approaching the cold-start median of 8. The one structure missed by both arms remains at $|n_{\mathrm{imp}}-n_{\mathrm{latt}}|\approx0.056$ regardless of initialization. This behavior points to a hard, plausibly multistable case (Sec.~\ref{sec:fixedpoint}), not a specific warm-start failure. 

The axis-aligned nature of this gap raises an obvious simpler baseline: interpolation in volume. We evaluated nearest-donor copying, inverse-distance-weighted averaging over four donors, and linear interpolation between donors bracketing the window [Fig.~\ref{fig:covgap}(b,c)]. Each was compared with the converged label using the per-structure $\Sigma(i\omega_n)$ metric of Fig.~\ref{fig:occconvergenceandsigpreds}(a).

The outcome separates the self-energy from the chemical potential. Nearest-donor copying gives a median self-energy error of 0.005~eV, three times smaller than the learned model's 0.018~eV. Yet the same donor misses $\mu$ by 0.55~eV, explaining why an apparently excellent self-energy transplant still enters the CSC loop slowly. Linear interpolation leaves the $\Sigma$ error at 0.005~eV but reduces the $\mu$ error to 0.066~eV, below the learned model's 0.125~eV. On a gap with one known coordinate and donors on both sides, hand-built interpolation is not merely competitive; for $\mu$, it is better.

The in-loop counts bear this out [Fig.~\ref{fig:covgap}(d)]. Volume interpolation reaches sustained convergence in a median of one iteration, compared with two for learning, seven for donor retrieval, and eight for cold starts. Eleven structures sustain the criterion under every viable arm; the twelfth defeats all initializations, including the cold start. Same-stack calibration replicas reproduce the archived learned-arm counts exactly.

What learning provides here is not superiority on the one-dimensional slice, but interpolation without a prescribed coordinate or deployment-time donor bookkeeping. A compositional or thermal gap has no obvious analogue of volume along which to construct this baseline. For that reason, we call the experiment a prospectively specified coverage-gap interpolation test, not a general out-of-distribution test. The model-side $1.2\times$ degradation and the in-loop $3.5\times$ iteration ratio express the same distinction between learned interpolation and single-donor retrieval.

\section{Application: correlated thermodynamics of iron at core conditions}
\label{sec:iron}

\subsection{A correlation-consistent interatomic potential}

We used the acceleration to label another 600 Fe configurations at core pressures and temperatures, making the data-generation campaign the method's first consumer. The resulting pool contains 1100 configurations with correlated energies and forces, of which 1052 are unique converged structures (Appendix~\ref{app:training}). For each new structure, the learned self-energy and Fermi level warm-started the CSC loop. These labels then trained the NequIP-based interatomic potentials used below~\cite{nequipTan2025,nequipBatzner2022}; training and deployment details are given in Appendix~\ref{app:mlip}.

On the archived 108-frame test split, the deployed models give an energy MAE of 69.2~meV/atom and a force-component MAE of 59.2~meV/\angstrom{}. The pooled force RMSE is 87.2~meV/\angstrom{}. We separately checked that the four coexistence points were assigned to the correct archived runs by remeasuring finite-difference pressures from their final frames. The resulting values, $307.9\pm0.4$, $322.9\pm0.4$, $337.2$, and $352.4\pm0.5$~GPa, reproduce the published pressures within 0.2~GPa. Thus each reported pressure is a measured stress, not the nominal directory label; the 337.2~GPa run is the state audited below.

Finite CTQMC statistics contribute only 0.1--0.5~meV/atom to the reference energies (Sec.~\ref{sec:fep}), far less than the model error~\cite{huangMachineLearningDiffusion2023}. At 5000~K, the 69.2~meV/atom test error is also below $k_BT\approx430$~meV. The force error is comparable to values reported for other finite-temperature and extreme-condition MLIPs used in melting studies~\cite{pmpDeng2021,pmpDeng2023,pmpFan2025,pmpPun2020,pmpTang2025,pmpWillman2022}. Nor is the result sensitive to the electronic label temperature: changing it from 5000 to 6000~K shifts energies by at most 6~meV/atom. None of these global metrics, however, directly gives an uncertainty in $T_m$. Melting depends on phase-resolved free energies, which we examine in Sec.~\ref{sec:fep}.

\subsection{Melting curve of hcp iron}

\begin{figure}[t]
    \centering
    \includegraphics[width=\linewidth]{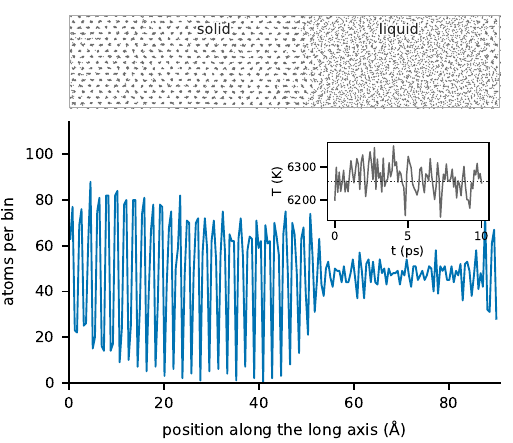}
    \caption{\textbf{Solid--liquid coexistence at the audited production state point.} Density profile along the long axis of the 9216-atom cell (quarter-wavelength bins) at the exact configuration used for the phase-density measurement of the free-energy audit ($T = 6256$~K, measured pressure $336.9\pm1.1$~GPa; Sec.~\ref{sec:fep}). Top: the atomic configuration, with the layered solid and disordered liquid regions labeled. Inset: temperature over the final 10~ps continuation segment (dotted line: 6256~K). The demodulation-derived phase blocks span 52.6 and 38.1~\AA{} (Appendix~\ref{app:coexistence}).}
    \label{fig:densityprofile300gpa}
\end{figure}

We determine melting temperatures by solid--liquid coexistence in the NVE ensemble, avoiding the hysteresis of single-phase superheating and supercooling~\cite{tpcMorris1994,aimcAlfe2009}. An elongated 9216-atom cell contains hcp and liquid regions: above $T_m$ the solid recedes, whereas below it the liquid crystallizes. Local bond-order analysis tracks the phase fractions~\cite{pfaLechner2008}. At the audited state, Fig.~\ref{fig:densityprofile300gpa} shows crystalline density oscillations, a smooth liquid profile, stable interfaces, and a steady coexistence temperature. After 8~ps of thermalization, trajectories run for 20--30~ps with a 0.5~fs timestep. Finite-difference stresses, including the kinetic term, are averaged over the final 5~ps; residual anisotropy of 2--3~GPa contributes a separate pressure uncertainty. The training data include liquid-like and hot solid configurations, and the coexistence trajectory remains stable for tens of picoseconds. Its stationary interface, smooth interfacial density, and integration-limited NVE drift support interpolation through the interface region (Appendix~\ref{app:coexistence}; Supplemental Notes~9 and~11).

\begin{figure}[t]
    \centering
    \includegraphics[width=\linewidth]{Tm.pdf}
    \caption{\textbf{Melting curve of iron at core pressures.} Melting curve from our solid--liquid coexistence simulations compared with experimental data~\cite{emcAnzellini2013,emcSinmyo2019,emcLi2020,emcKraus2022,emcBalugani2024} and prior ab initio predictions~\cite{aimcAlfe2009,aimcGonzalez2023,aimcSun2022,aimcSun2023,aimcBelonoshko2021,aimcStixrude2014,aimcAlfe2002,aimcBelonosko2000,aimcWu2024,aimcSola2009}. Statistical uncertainties of our points are smaller than the symbols. The plotted point is the raw result of the production protocol; no DFT+DMFT temperature correction is assigned (Sec.~\ref{sec:fep}).}
  \label{fig:femeltcurve}
\end{figure}

Our melting results, spanning 307--352~GPa (Fig.~\ref{fig:femeltcurve}), are well described by a Simon--Glatzel relation~\cite{simonglatz},
\begin{equation}
    T_m = T_0\left(\frac{P-P_0}{a} + 1\right)^{1/c},
\end{equation}
with $T_0=6167$~K, $P_0=323.1$~GPa, $a=298.73$~GPa, and $c=2.43$. At 330~GPa, the inner-core-boundary pressure, we obtain $T_m=6225\pm42\,\mathrm{K}_{\mathrm{stat}}$; this uncertainty reflects fluctuations within the coexistence MD. Model fidelity is examined separately below. Our value agrees with the laser-heated diamond-anvil result of $6230\pm500$~K from Anzellini et al., extrapolated upward from roughly 200~GPa~\cite{emcAnzellini2013}. It is also consistent with Kraus et al., who extrapolate downward from roughly 600~GPa~\cite{emcKraus2022}, and Balugani et al.'s $6202\pm514$~K shock-XAS bound~\cite{emcBalugani2024}. The result lies above Sinmyo et al.'s resistance-heated value of $5500\pm220$~K~\cite{emcSinmyo2019}. Static experiments can differ systematically because they use different melting criteria~\cite{emcAnzellini2013}.

\subsection{Model error at the free-energy level}
\label{sec:fep}

Because melting is set by equal solid and liquid chemical potentials, the global MLIP energy MAE is not the error that controls $T_m$. What matters is the phase-resolved free-energy separation between the MLIP and DFT+DMFT surfaces~\cite{fepJung2023,fepSun2018}, which we evaluate by free-energy perturbation. With $\Delta u=u_{\mathrm{DMFT}}-u_{\mathrm{MLIP}}$ denoting the per-atom energy difference and $N$ the cell size, the correction for phase $x\in\{\mathrm{solid},\mathrm{liquid}\}$ is
\begin{equation}
\Delta F_x = -\frac{1}{\beta N}\ln\left\langle e^{-\beta N \Delta u}\right\rangle_x
\;\approx\; \langle \Delta u\rangle_x \;-\; \frac{\beta N}{2}\,\sigma_x^2(\Delta u),
\label{eq:fep}
\end{equation}
A preliminary survey used a $k$-means solid-like/liquid-like partition of the held-out training pool as a proxy for the two phases. Retained in Supplemental Note~8, that survey motivated the dedicated four-atom ensembles evaluated here at the measured coexistence state.

Whether the variance term is a bound or a measurement depends on how much of the residual width is CTQMC label noise. We measured this directly with independent-seed repeats of identical warm restarts (three structures spanning the force-magnitude range; seeds injected because the solver's default seeding is deterministic). Seed-to-seed energy differences are only \textbf{0.1--0.5~meV/atom} at matched iterations, a factor of $10^2$--$10^3$ smaller in variance than the 86--135~meV/atom residual widths. The residual spread is therefore model error, essentially none of it label noise, and the second-cumulant term is real rather than noise-inflated.

At 6256~K we evaluated 25 configurations at each production-derived specific volume, with all 56 DFT+DMFT calculations, including six CTQMC replicas, satisfying sustained convergence. We call the correction difference between the 7.1388 and 7.0122~\AA$^3$/atom sets $\Delta\Delta A_V$, reserving solid--liquid notation for phase-valid ensembles. The direct exponential estimator gives $\Delta\Delta A_V=448$~meV/atom, with a paired-bootstrap 95\% confidence interval of 423--473~meV/atom. Effective-sample-size fractions are 0.89 and 0.91, and the second-cumulant shift is only $-9$~meV/atom, indicating a statistically well-conditioned estimator.

This numerical convergence does not establish a melting correction, and we do not report one. A layering order parameter gives $|S| = 0.92$--0.95 for both sets; a four-atom periodic cell cannot realize a liquid. Their potential-energy difference follows the crystal compression energy rather than any liquid configurational contribution. The two sets therefore sample the same crystalline basin at two specific volumes, not distinct solid and liquid ensembles. Dividing $\Delta\Delta A_V$ by an entropy of melting would therefore manufacture a temperature shift from a quantity that is not a solid--liquid free-energy difference. We report the audit as a density-conditioned calibration measurement: the deployed potential's DFT+DMFT correction changes by up to 448~meV/atom between the two coexistence densities. The raw $T_m = 6225 \pm 42$~K$_{\mathrm{stat}}$ stands as the result of the production protocol, with no DFT+DMFT temperature correction assigned.

\section{Discussion}
\label{sec:discussion}

No single initializer wins across all regimes. In a dense, same-state archive, the local redundancy of the CSC fixed point makes a nearby converged solution difficult to beat. Retrieval loses that advantage at the coverage boundary, where the available donors become progressively less representative of the target. This boundary is not a fixed property of a material: it moves as a campaign grows, while new compositions and state points remain outside the validated domain.

The pre-registered window puts numbers on this transition. Median convergence takes 2 iterations with learning, 7 with retrieval, and 8 from a cold start (Sec.~\ref{sec:ood}). Thus retrieval gives up most of its benefit when forced across the gap, whereas the learned functional retains nearly all of its in-distribution acceleration. The solver still protects the final observable by converting initialization error into extra iterations, except that rugged landscapes admit more than one stationary branch. There, charge convergence alone is insufficient and should be accompanied by an energy-stationarity check.

Inside dense coverage, learning offers reliability rather than greater median speed: it removes the observed failure tail and dispenses with an online archive at deployment. A practical workflow could use retrieval when a close, state-compatible donor exists and fall back to learning when it does not. More generally, claims for learned initialization should be tested against the strongest data-reuse baseline available for the solver in question.

For a campaign still in progress, the coverage boundary is the usual operating point rather than an edge case. New compute is spent on the next compression, distortion batch, or sparsely sampled volume, precisely where the archive is weakest. In the held-out window, learning removes six of the eight cold-start iterations, whereas a donor transplant removes only one. This experiment also changes how the production history in Sec.~\ref{sec:iron} should be read. Without it, the earlier speedups could be attributed to a dense archive that makes lookup and learning equally effective. Retrieval fails exactly where learning does not, showing that coverage alone cannot explain the production acceleration.

Giuli et al. attack a complementary part of the problem by compressing the variational space of a ghost-Gutzwiller embedding solver with a linear foundation model~\cite{giuli2026linear}. Their acceleration applies to the embedding-Hamiltonian step at zero temperature and transfers across lattice models. Our model instead starts from atomic structure, operates at finite temperature within full charge self-consistency, and retains the numerically exact CTQMC impurity solver. The resulting fixed points provide forces and, through the MLIP, thermodynamics. These approaches could be composed: impurity-solver compression would multiply the per-iteration savings of a warm-started CSC loop. In both cases, learning is safest where the underlying theory subsequently refines or verifies the learned object.

The evidence has clear limits. Our four-arm comparison contains only 24 Fe structures, so the retrieval failure tail is supported by single-digit counts. The nearest-sibling arm is also an oracle bound, since some donors occur later than their targets in campaign chronology. Fingerprint-nearest retrieval better represents what was operationally available. Learning is not infrastructure-free either: it needs the same converged corpus that retrieval uses as an archive. Their practical difference is therefore coverage and reliability, not the amortized cost of that corpus (Sec.~\ref{sec:accel}).

Two model limitations remain equally concrete. The orbital readouts are consistent within the projector frame but not covariant under arbitrary rigid rotations, and causality is validated rather than enforced (Sec.~\ref{sec:method}). The Fe application should also be read as a benchmark, not a discovery, because careful DFT calculations already overlap current experimental bounds~\cite{fepSun2018,aimcAlfe2009}. Here a dynamical treatment of Fe $3d$ correlations reaches that same melting range without adjustment, using interaction parameters adopted rather than fitted from earlier work~\cite{ieccPourovskii2017,chsGavriliuk2023,chsHo2024}. DMFT carries the calculation through large-scale sampling while treating instantaneous local moments and quasiparticle renormalization on the same footing. At core conditions those moments remain sizable, and their entropy differs between phases~\cite{PhysRevB.87.014405,ieccPourovskii2017,ieccVekilova2015}. Isolating their contribution will require the controlled comparison now made possible: DFT and DFT+DMFT potentials trained on identical configurations.

The crystallizing phase of the inner core is the most immediate target. Bcc iron has now been observed near melting above 200~GPa~\cite{emcKonopkova2025}, where competing free energies differ by only tens of meV/atom. Spin-fluctuation entropy may be decisive at this scale, requiring correlated free energies and large-cell sampling in the same calculation. Other applications include FeO and Fe alloys at core--mantle-boundary conditions, electron--electron-scattering transport, and finite-temperature phase diagrams of correlated oxides. Because the initializer already lands near self-consistency, it may eventually support uncertainty-controlled refinement, with full CTQMC invoked selectively rather than for every configuration. Such a mode could extend the speedup beyond the present factor of four without giving up solver verification.

\appendix

\section{DFT+DMFT reference calculations}
\label{app:dmft}

The DFT part of the CSC loop uses the local-density approximation with the augmented-plane-wave plus local-orbital method as implemented in WIEN2k~\cite{wien2kblaha2020}. For Fe we use a muffin-tin radius of 1.6~bohr, RKMax $= 7.5$, core--valence separation $-7.5$~Ry, a k-mesh scaled as 2000/(number of Fe atoms), and charge convergence of $10^{-5}$. FeO uses muffin-tin radii of 1.6 (Fe) and 1.4~bohr (O) with the same RKMax, separation, k-mesh scheme, and convergence; NiO uses 1.8 (Ni) and 1.5~bohr (O), RKMax $= 7.5$, $-6$~Ry, 1000 k-points, and charge convergence of $10^{-3}$. The DMFT part is handled by the eDMFT package~\cite{edmftfHaule2010,edmftfHaule2015}, at 5000~K for Fe and FeO and 611~K for NiO. Density--density interactions $U = 5.0$ (Fe), 10.0 (FeO), 8.0~eV (NiO) and Hund's couplings $J = 0.93$, 1.0, 0.9~eV are adopted from previous studies~\cite{ieccPourovskii2017,chsGavriliuk2023,chsHo2024} and held constant. Projector windows span $-10$ to $10$~eV (NiO, FeO) and $-12.2$ to $4$~eV (Fe). The impurity problem is solved with the CTQMC solver of eDMFT~\cite{edmftfHaule2007} using $10^8$ Monte Carlo steps per processor on 32 processors (Fe, FeO) and $10^7$ steps per processor on 32 processors (NiO), verified against the archived per-run inputs. Calculations stop at the first crossing of $|n_{\mathrm{imp}} - n_{\mathrm{lat}}| \le 0.01$ (the historical production rule); because the CTQMC occupancy noise floor is ${\approx}0.02$, all iteration counts and convergence claims in this work instead use the stricter sustained criterion of Sec.~\ref{sec:accel} ($|n_{\mathrm{imp}} - n_{\mathrm{lat}}| < 0.05$ maintained to the end of the run), which is immune to transient first crossings. Forces derive from the Luttinger--Ward functional~\cite{edmftfHaule2016}, double counting is nominal~\cite{edmftfHaule2010,dcHaule2014}, spin--orbit coupling is neglected for the $3d$ elements, and all systems are paramagnetic with local-moment fluctuations included at the DMFT level.

\section{Training-set generation}
\label{app:training}

For Fe and FeO, configurations were sampled from finite-temperature ab initio MD using VASP~\cite{vaspKresse1996,vaspKresse1999}, with snapshots selected by $k$-medoids clustering in structural-fingerprint space~\cite{fpZhu2016} to maximize the diversity of local environments. For Fe, 2.5-ps trajectories (0.5~fs timestep) of 4--8-atom cells were run for hcp, bcc, and fcc at 300~GPa and 6000~K, yielding ${\sim}500$ selected configurations. Two enhancement sets were added: 200 configurations from NVT volume-scan MD on hcp Fe at 5000~K spanning 6.5--8.3~\AA$^3$/atom, and 400 solid- and liquid-like configurations selected from M3GNet-driven~\cite{m3gChen2022} NPT MD at 10\,000~K targeted at 330~GPa. For these additional structures the learned functional supplied the warm start, bringing the Fe pool to 1100 labeled configurations, of which 1052 unique converged structures were retained. For FeO, the B2 phase~\cite{feoOzawa2011} was sampled at 10\,000~K in NVT (500 configurations); for NiO, we use the 1500-configuration perturbation dataset of our previous study~\cite{Rao2024}. Full details in Supplemental Notes~1--3.

\section{Network architecture and training}
\label{app:network}

Atoms are nodes of a periodic graph with edges within a radial cutoff; edge features are spherical harmonics combined with learned radial bases (10 Gaussian basis functions, two 64-neuron radial layers, SiLU activations). We use two interaction layers, embedding dimension 64, maximum irreducible-representation order 2, and gated nonlinearities preserving equivariance. Orbital-specific readout networks (one instantiation per orbital, shared architecture, independent weights) output the Legendre coefficients and $\Sigma(\infty)$ for each symmetrically inequivalent correlated atom; a pooled scalar readout yields $E_f$. Predicted components are reconstructed to $\Sigma^{\mathrm{ML}}(i\omega_n)$ with TRIQS~\cite{Parcollet2015,Boehnke2011}. Models were trained for 20 epochs with AdamW~\cite{adamwLoshchilov2019} at initial learning rate 0.01 with validation-adaptive scheduling and smooth-L1 loss. Details in Supplemental Note~1.

\section{Machine-learned interatomic potential}
\label{app:mlip}

The Fe MLIPs use NequIP~\cite{nequipTan2025,nequipBatzner2022}: cutoff 5~\angstrom, four interaction blocks with 32 features, maximum irrep order 2, two 64-neuron radial layers, 8 radial Bessel functions with polynomial cutoff 6, and a ZBL pair potential~\cite{zblZiegler1985} for close-range repulsion. The 1052 unique configurations were split 80/10/10 into training/validation/test. Evaluated directly on the archived test split with the deployed models, the energy MAE is 69.2~meV/atom and the force-component MAE 59.2~meV/\angstrom{} (RMSE 87.2); the solid-phase equation of state is validated against experiment (Supplemental Note~5). Parity plots are in Supplemental Note~4, and a pressure--volume validation against experimental equation-of-state data is in Supplemental Note~5. The reported melting point is a result of this production protocol, with no DFT+DMFT correction applied.

\section{Solid--liquid coexistence protocol}
\label{app:coexistence}

Coexistence simulations follow the NVE protocol~\cite{tpcMorris1994,aimcAlfe2009} in the atomic simulation environment~\cite{aseLarsen2017}. A 9216-atom, nearly orthorhombic supercell built from hcp Fe (lattice parameters from experiment~\cite{felpHirao2022}, optimized by finite-difference stress at each pressure) is thermalized near the estimated melting point in NVT with Langevin dynamics. Half the cell is frozen while the other half is melted at elevated temperature; the liquid is then re-thermalized. The full system is evolved in NVE for 20--30~ps with a 0.5~fs timestep. Phase fractions are monitored by local bond-order analysis~\cite{pfaLechner2008} and confirmed visually. Pressure is computed by finite differences under cell deformations, including the kinetic contribution through the virial, and averaged with temperature over the final 5~ps. Melting-point data, phase-fraction traces, and stress-anisotropy analysis in Supplemental Notes~6 and~11.

\section{Retrieval baselines and the four-arm protocol}
\label{app:baselines}

The comparison uses 24 structures, split evenly between the strongly distorted production batch and the near-ideal volume scan. Seeded random sampling selected them from converged warm-started runs, and the list was recorded before any baseline calculation. WIEN2k initialization, DMFT parameters, and iteration budgets are identical across the four arms; only the initial $\{\Sigma,E_f\}$ changes.

Cold starts set $\Sigma_{\mathrm{dyn}}=0$ and initialize $\Sigma(\infty)$ at the nominal double counting, whereas learned starts use the stored production predictions. Fingerprint-nearest retrieval (nn) copies the converged self-energy and chemical potential of the closest cold-labeled structure~\cite{fpZhu2016}. Gaussian-overlap-matrix fingerprints and Hungarian atom assignment define proximity, but donor eligibility also requires the same impurity structure, thermodynamic state, and Matsubara grid. That last restriction matters: our initial hash-level analysis silently paired 2000~K donors with 5000~K targets. Nearest-sibling retrieval (adj) instead copies the closest structure from the same batch, providing an oracle bound because it ignores campaign chronology.

Warm replicas make the new calculations comparable to the two-year-old archive, reproducing $|n_{\mathrm{imp}}-n_{\mathrm{lat}}|$ to three decimal places at each iteration. Energy comparisons apply the $+6.6\pm5$~meV/cell residual stack offset measured from three converged-region replicas. Archive tails are truncated to the matched iteration budgets.

The component-matrix runs of Sec.~\ref{sec:accel} reuse this protocol unchanged: for each eligible structure the two mixed arms were staged from byte-verified inputs (donor $\Sigma$ files and predicted $\mu_0$ from the archived production iterates), cold-started Wien2k densities, and the identical convergence accounting. Donor eligibility required a converged donor with the production orbital setup; the single candidate whose only donor used an earlier single-impurity configuration (a different correlated subspace) was excluded from all four arms.

\section{The pre-registered coverage-gap (OOD) protocol}
\label{app:ood}

The training corpus comprises the 350 cold-labeled configurations of the four Fe production label sets. The holdout is the volume window $V/\mathrm{atom} \in [7.9, 9.8]$~\AA$^3$ (39 structures, 11\%), fixed and written to a manifest before any training; the training driver verifies the manifest and refuses to run if it changes. Two model sets are trained with identical hyperparameters (those of Appendix~\ref{app:network}): the OOD model on the 311 structures outside the window, and a full-corpus control that separates retraining quality from coverage effects. For the retrieval comparison, donor eligibility excludes the window. Model-side evaluation scores $\{\Sigma_\ell, \Sigma(\infty), E_f\}$ predictions of both models against the held-out structures' true converged labels; in-loop verification runs DFT+DMFT on a seeded 12-structure subset with learned-OOD, nearest-eligible-donor, and volume-interpolation initializations, against the structures' archived cold trajectories; all measured results appear in Sec.~\ref{sec:ood}. The volume-interpolation arm ran after a cluster migration; two calibration replicas of learned-arm runs, executed on the new stack from identical inputs, reproduced their archived iteration counts exactly, validating cross-stack comparability of the counts. One donor-initialized run lost its first attempt to a node failure at iteration 3 and was re-run from the identical donor state; the repeat converged at iteration 8 and is the value reported. Runs are capped at 13 iterations, matching the validation-campaign budget, and convergence is the sustained criterion throughout; we label the first impurity solve iteration 1.

\section{Label-noise measurement}
\label{app:noise}

Independent-seed repeats of identical self-warm restarts (each run restarted from its own converged state; complete seed pairs for two structures spanning the solid-like and intermediate force range, plus a single-seed run of a liquid-like structure whose within-run behavior is consistent) isolate the CTQMC statistical noise on energy labels. Because the solver's default seeding is deterministic ($\mathrm{seed} = 10\,\mathrm{rank} + \mathrm{iseed}$ with $\mathrm{iseed}=0$), distinct \texttt{iseed} values were injected per repeat; iteration-1 energies are bit-identical across seeds (a deterministic pre-solve value, confirming the injection), and subsequent matched-iteration differences give $\sigma_E \approx 0.1$--$0.5$~meV/atom at production statistics ($10^8$ steps/process $\times$ 32). Iteration-to-iteration drift within a settling run is separately ${\sim}6$~meV/atom and decays with settling.

\section{Free-energy-perturbation analysis}
\label{app:fep}

The FEP analysis uses the 106 four-atom Fe configurations of the held-out pool. Signed residuals $\delta E_i = E_i^{\mathrm{MLIP}} - E_i^{\mathrm{DFT+DMFT}}$ (so $\Delta u = -\delta E$) are partitioned into solid-like and liquid-like clusters by $k$-means ($k=2$) in the feature space of per-configuration force RMS and energy above a cold lower-envelope at the same atomic volume; threshold partitions on either feature serve as robustness checks. For each partition and each trained architecture (production NequIP and an Allegro-based fine-tuned model), both cumulant terms of Eq.~\eqref{eq:fep} are evaluated at $T_m = 6225$~K with $N=4$ and propagated through $\Delta S_m = L_{\mathrm{melt}}/T_m$ with $L_{\mathrm{melt}} = 0.5$--0.6~eV/atom (conservative bracket 0.4--0.8)~\cite{aimcAlfe2002,aimcAlfe2009}. Because the residual variance contains the CTQMC label-noise floor, and because four-atom statistics only weakly constrain the spatial correlation of model errors, the second-order term is reported as an upper bound; the complete table is in Supplemental Note~8. This survey is superseded by the coexistence-ensemble audit below and is retained as motivation.

The definitive analysis evaluates both deployed potentials and fully charge-self-consistent DFT+DMFT on 25 four-atom configurations at each production-derived specific volume, generated by Langevin dynamics with the production calculator at the measured coexistence temperature (the liquid-volume set prepared by melt--quench--equilibrate before sampling); six CTQMC replicas give 56 converged calculations in total. The direct exponential estimator gives a density-conditioned differential $\Delta\Delta A_V = 448$~meV/atom (paired-bootstrap 95\% confidence interval 423--473~meV/atom; bootstrap standard deviation 12.7~meV/atom), with effective-sample-size fractions 0.89 and 0.91. Both four-atom sets remain crystalline under the layering diagnostic ($|S| = 0.92$--0.95); we therefore identify them by their production-derived specific volumes rather than as solid and liquid phases, and interpret the result as an audit of density-dependent energy calibration rather than as $\Delta F_l - \Delta F_s$ or a melting-temperature correction. As an out-of-sample stability check on the dominant term, a disjoint, independently seeded 180-frame ensemble at each specific volume reproduces it within statistical compatibility. The bridge magnitude is thus a property of the ensembles, not of the 25 audited configurations. This check does not address the four-atom phase limitation. The six replica pairs bound the CTQMC contribution to the per-cell scatter at the ${\sim}50$~meV level, an order of magnitude below the measured widths.

\section*{Data availability}
The SigML package used for prediction of the self-energies and Fermi levels is openly available at \url{https://github.com/Rutgers-ZRG/SigML}. All other data supporting the findings of this study, including the DFT+DMFT-labeled training configurations, the trained interatomic-potential parameters, and the solid--liquid coexistence trajectories, are available from the corresponding author upon reasonable request.

\begin{acknowledgments}
The authors acknowledge the Office of Advanced Research Computing (OARC) at Rutgers University for providing access to the Amarel cluster and associated research computing resources. This work was supported by the start-up funds of the Office of the Dean of the School of Arts and Sciences--Newark (SASN), Rutgers University--Newark.
\end{acknowledgments}

\bibliography{refs}

\begin{thebibliography}{72}%
\makeatletter
\providecommand \@ifxundefined [1]{%
 \@ifx{#1\undefined}
}%
\providecommand \@ifnum [1]{%
 \ifnum #1\expandafter \@firstoftwo
 \else \expandafter \@secondoftwo
 \fi
}%
\providecommand \@ifx [1]{%
 \ifx #1\expandafter \@firstoftwo
 \else \expandafter \@secondoftwo
 \fi
}%
\providecommand \natexlab [1]{#1}%
\providecommand \enquote  [1]{``#1''}%
\providecommand \bibnamefont  [1]{#1}%
\providecommand \bibfnamefont [1]{#1}%
\providecommand \citenamefont [1]{#1}%
\providecommand \href@noop [0]{\@secondoftwo}%
\providecommand \href [0]{\begingroup \@sanitize@url \@href}%
\providecommand \@href[1]{\@@startlink{#1}\@@href}%
\providecommand \@@href[1]{\endgroup#1\@@endlink}%
\providecommand \@sanitize@url [0]{\catcode `\\12\catcode `\$12\catcode
  `\&12\catcode `\#12\catcode `\^12\catcode `\_12\catcode `\%12\relax}%
\providecommand \@@startlink[1]{}%
\providecommand \@@endlink[0]{}%
\providecommand \url  [0]{\begingroup\@sanitize@url \@url }%
\providecommand \@url [1]{\endgroup\@href {#1}{\urlprefix }}%
\providecommand \urlprefix  [0]{URL }%
\providecommand \Eprint [0]{\href }%
\providecommand \doibase [0]{https://doi.org/}%
\providecommand \selectlanguage [0]{\@gobble}%
\providecommand \bibinfo  [0]{\@secondoftwo}%
\providecommand \bibfield  [0]{\@secondoftwo}%
\providecommand \translation [1]{[#1]}%
\providecommand \BibitemOpen [0]{}%
\providecommand \bibitemStop [0]{}%
\providecommand \bibitemNoStop [0]{.\EOS\space}%
\providecommand \EOS [0]{\spacefactor3000\relax}%
\providecommand \BibitemShut  [1]{\csname bibitem#1\endcsname}%
\let\auto@bib@innerbib\@empty
\bibitem [{\citenamefont {Kotliar}\ \emph {et~al.}(2006)\citenamefont
  {Kotliar}, \citenamefont {Savrasov}, \citenamefont {Haule}, \citenamefont
  {Oudovenko}, \citenamefont {Parcollet},\ and\ \citenamefont
  {Marianetti}}]{dmftKotliar2006}%
  \BibitemOpen
  \bibfield  {author} {\bibinfo {author} {\bibfnamefont {G.}~\bibnamefont
  {Kotliar}}, \bibinfo {author} {\bibfnamefont {S.~Y.}\ \bibnamefont
  {Savrasov}}, \bibinfo {author} {\bibfnamefont {K.}~\bibnamefont {Haule}},
  \bibinfo {author} {\bibfnamefont {V.~S.}\ \bibnamefont {Oudovenko}}, \bibinfo
  {author} {\bibfnamefont {O.}~\bibnamefont {Parcollet}},\ and\ \bibinfo
  {author} {\bibfnamefont {C.~A.}\ \bibnamefont {Marianetti}},\ }\bibfield
  {title} {\bibinfo {title} {Electronic structure calculations with dynamical
  mean-field theory},\ }\href {https://doi.org/10.1103/RevModPhys.78.865}
  {\bibfield  {journal} {\bibinfo  {journal} {Rev. Mod. Phys.}\ }\textbf
  {\bibinfo {volume} {78}},\ \bibinfo {pages} {865} (\bibinfo {year}
  {2006})}\BibitemShut {NoStop}%
\bibitem [{\citenamefont {Abrahams}\ and\ \citenamefont
  {Kotliar}(1996)}]{mitAbrahams1996}%
  \BibitemOpen
  \bibfield  {author} {\bibinfo {author} {\bibfnamefont {E.}~\bibnamefont
  {Abrahams}}\ and\ \bibinfo {author} {\bibfnamefont {G.}~\bibnamefont
  {Kotliar}},\ }\bibfield  {title} {\bibinfo {title} {The metal-insulator
  transition in correlated disordered systems},\ }\href
  {https://doi.org/10.1126/science.274.5294.1853} {\bibfield  {journal}
  {\bibinfo  {journal} {Science}\ }\textbf {\bibinfo {volume} {274}},\ \bibinfo
  {pages} {1853} (\bibinfo {year} {1996})}\BibitemShut {NoStop}%
\bibitem [{\citenamefont {Imada}\ \emph {et~al.}(1998)\citenamefont {Imada},
  \citenamefont {Fujimori},\ and\ \citenamefont {Tokura}}]{mitMasatoshi1998}%
  \BibitemOpen
  \bibfield  {author} {\bibinfo {author} {\bibfnamefont {M.}~\bibnamefont
  {Imada}}, \bibinfo {author} {\bibfnamefont {A.}~\bibnamefont {Fujimori}},\
  and\ \bibinfo {author} {\bibfnamefont {Y.}~\bibnamefont {Tokura}},\
  }\bibfield  {title} {\bibinfo {title} {Metal-insulator transitions},\ }\href
  {https://doi.org/10.1103/RevModPhys.70.1039} {\bibfield  {journal} {\bibinfo
  {journal} {Rev. Mod. Phys.}\ }\textbf {\bibinfo {volume} {70}},\ \bibinfo
  {pages} {1039} (\bibinfo {year} {1998})}\BibitemShut {NoStop}%
\bibitem [{\citenamefont {Anisimov}\ \emph {et~al.}(1997)\citenamefont
  {Anisimov}, \citenamefont {Aryasetiawan},\ and\ \citenamefont
  {Lichtenstein}}]{tmoAnisimov1997}%
  \BibitemOpen
  \bibfield  {author} {\bibinfo {author} {\bibfnamefont {V.~I.}\ \bibnamefont
  {Anisimov}}, \bibinfo {author} {\bibfnamefont {F.}~\bibnamefont
  {Aryasetiawan}},\ and\ \bibinfo {author} {\bibfnamefont {A.~I.}\ \bibnamefont
  {Lichtenstein}},\ }\bibfield  {title} {\bibinfo {title} {First-principles
  calculations of the electronic structure and spectra of strongly correlated
  systems: the \text{LDA+U} method},\ }\href
  {https://doi.org/10.1088/0953-8984/9/4/002} {\bibfield  {journal} {\bibinfo
  {journal} {J. Phys. Condens. Matter}\ }\textbf {\bibinfo {volume} {9}},\
  \bibinfo {pages} {767} (\bibinfo {year} {1997})}\BibitemShut {NoStop}%
\bibitem [{\citenamefont {Tokura}\ and\ \citenamefont
  {Nagaosa}(2000)}]{tmoTokura2000}%
  \BibitemOpen
  \bibfield  {author} {\bibinfo {author} {\bibfnamefont {Y.}~\bibnamefont
  {Tokura}}\ and\ \bibinfo {author} {\bibfnamefont {N.}~\bibnamefont
  {Nagaosa}},\ }\bibfield  {title} {\bibinfo {title} {Orbital physics in
  transition-metal oxides},\ }\href
  {https://doi.org/10.1126/science.288.5465.462} {\bibfield  {journal}
  {\bibinfo  {journal} {Science}\ }\textbf {\bibinfo {volume} {288}},\ \bibinfo
  {pages} {462} (\bibinfo {year} {2000})}\BibitemShut {NoStop}%
\bibitem [{\citenamefont {Paul}\ and\ \citenamefont
  {Birol}(2019)}]{dmftArpita2019}%
  \BibitemOpen
  \bibfield  {author} {\bibinfo {author} {\bibfnamefont {A.}~\bibnamefont
  {Paul}}\ and\ \bibinfo {author} {\bibfnamefont {T.}~\bibnamefont {Birol}},\
  }\bibfield  {title} {\bibinfo {title} {Applications of \text{DFT} +
  \text{DMFT} in materials science},\ }\href
  {https://doi.org/https://doi.org/10.1146/annurev-matsci-070218-121825}
  {\bibfield  {journal} {\bibinfo  {journal} {Annu. Rev. Mater. Res.}\ }\textbf
  {\bibinfo {volume} {49}},\ \bibinfo {pages} {31} (\bibinfo {year}
  {2019})}\BibitemShut {NoStop}%
\bibitem [{\citenamefont {Arsenault}\ \emph {et~al.}(2014)\citenamefont
  {Arsenault}, \citenamefont {Lopez-Bezanilla}, \citenamefont {von
  Lilienfeld},\ and\ \citenamefont {Millis}}]{mlgfArsenault2014}%
  \BibitemOpen
  \bibfield  {author} {\bibinfo {author} {\bibfnamefont {L.-F.}\ \bibnamefont
  {Arsenault}}, \bibinfo {author} {\bibfnamefont {A.}~\bibnamefont
  {Lopez-Bezanilla}}, \bibinfo {author} {\bibfnamefont {O.~A.}\ \bibnamefont
  {von Lilienfeld}},\ and\ \bibinfo {author} {\bibfnamefont {A.~J.}\
  \bibnamefont {Millis}},\ }\bibfield  {title} {\bibinfo {title} {Machine
  learning for many-body physics: The case of the anderson impurity model},\
  }\href {https://doi.org/10.1103/PhysRevB.90.155136} {\bibfield  {journal}
  {\bibinfo  {journal} {Phys. Rev. B}\ }\textbf {\bibinfo {volume} {90}},\
  \bibinfo {pages} {155136} (\bibinfo {year} {2014})}\BibitemShut {NoStop}%
\bibitem [{\citenamefont {Sturm}\ \emph {et~al.}(2021)\citenamefont {Sturm},
  \citenamefont {Carbone}, \citenamefont {Lu}, \citenamefont {Weichselbaum},\
  and\ \citenamefont {Konik}}]{mlgfSturm2021}%
  \BibitemOpen
  \bibfield  {author} {\bibinfo {author} {\bibfnamefont {E.~J.}\ \bibnamefont
  {Sturm}}, \bibinfo {author} {\bibfnamefont {M.~R.}\ \bibnamefont {Carbone}},
  \bibinfo {author} {\bibfnamefont {D.}~\bibnamefont {Lu}}, \bibinfo {author}
  {\bibfnamefont {A.}~\bibnamefont {Weichselbaum}},\ and\ \bibinfo {author}
  {\bibfnamefont {R.~M.}\ \bibnamefont {Konik}},\ }\bibfield  {title} {\bibinfo
  {title} {Predicting impurity spectral functions using machine learning},\
  }\href {https://doi.org/10.1103/PhysRevB.103.245118} {\bibfield  {journal}
  {\bibinfo  {journal} {Phys. Rev. B}\ }\textbf {\bibinfo {volume} {103}},\
  \bibinfo {pages} {245118} (\bibinfo {year} {2021})}\BibitemShut {NoStop}%
\bibitem [{\citenamefont {Dong}\ \emph {et~al.}(2024)\citenamefont {Dong},
  \citenamefont {Gull},\ and\ \citenamefont {Wang}}]{mlgfDong2024}%
  \BibitemOpen
  \bibfield  {author} {\bibinfo {author} {\bibfnamefont {X.}~\bibnamefont
  {Dong}}, \bibinfo {author} {\bibfnamefont {E.}~\bibnamefont {Gull}},\ and\
  \bibinfo {author} {\bibfnamefont {L.}~\bibnamefont {Wang}},\ }\bibfield
  {title} {\bibinfo {title} {Equivariant neural network for green's functions
  of molecules and materials},\ }\href
  {https://doi.org/10.1103/PhysRevB.109.075112} {\bibfield  {journal} {\bibinfo
   {journal} {Phys. Rev. B}\ }\textbf {\bibinfo {volume} {109}},\ \bibinfo
  {pages} {075112} (\bibinfo {year} {2024})}\BibitemShut {NoStop}%
\bibitem [{\citenamefont {Mitra}\ and\ \citenamefont
  {Banerjee}(2025)}]{mlgfMitra2025}%
  \BibitemOpen
  \bibfield  {author} {\bibinfo {author} {\bibfnamefont {P.}~\bibnamefont
  {Mitra}}\ and\ \bibinfo {author} {\bibfnamefont {H.}~\bibnamefont
  {Banerjee}},\ }\bibfield  {title} {\bibinfo {title} {Deep learning-based
  prediction of self-energies from *ab initio* dynamical mean-field theory for
  real materials with minimal data sets},\ }\bibfield  {journal} {\bibinfo
  {journal} {ChemRxiv}\ }\href {https://doi.org/10.26434/chemrxiv-2025-dp7rd}
  {10.26434/chemrxiv-2025-dp7rd} (\bibinfo {year} {2025})\BibitemShut {NoStop}%
\bibitem [{\citenamefont {Valenti}\ \emph {et~al.}(2026)\citenamefont
  {Valenti}, \citenamefont {Park}, \citenamefont {Georges}, \citenamefont
  {Millis},\ and\ \citenamefont {Parcollet}}]{valenti2026neural}%
  \BibitemOpen
  \bibfield  {author} {\bibinfo {author} {\bibfnamefont {A.}~\bibnamefont
  {Valenti}}, \bibinfo {author} {\bibfnamefont {I.}~\bibnamefont {Park}},
  \bibinfo {author} {\bibfnamefont {A.}~\bibnamefont {Georges}}, \bibinfo
  {author} {\bibfnamefont {A.~J.}\ \bibnamefont {Millis}},\ and\ \bibinfo
  {author} {\bibfnamefont {O.}~\bibnamefont {Parcollet}},\ }\bibfield  {title}
  {\bibinfo {title} {Neural-network quantum embedding solvers for correlated
  materials},\ }\href@noop {} {\bibfield  {journal} {\bibinfo  {journal} {arXiv
  preprint arXiv:2603.15741}\ } (\bibinfo {year} {2026})}\BibitemShut {NoStop}%
\bibitem [{\citenamefont {Nain}\ \emph {et~al.}(2026)\citenamefont {Nain},
  \citenamefont {Dee}, \citenamefont {Barros}, \citenamefont {Johnston},\ and\
  \citenamefont {Maier}}]{nain2026neural}%
  \BibitemOpen
  \bibfield  {author} {\bibinfo {author} {\bibfnamefont {R.}~\bibnamefont
  {Nain}}, \bibinfo {author} {\bibfnamefont {P.~M.}\ \bibnamefont {Dee}},
  \bibinfo {author} {\bibfnamefont {K.}~\bibnamefont {Barros}}, \bibinfo
  {author} {\bibfnamefont {S.}~\bibnamefont {Johnston}},\ and\ \bibinfo
  {author} {\bibfnamefont {T.~A.}\ \bibnamefont {Maier}},\ }\bibfield  {title}
  {\bibinfo {title} {Neural network as low-cost surrogates for impurity solvers
  in quantum embedding methods},\ }\href@noop {} {\bibfield  {journal}
  {\bibinfo  {journal} {arXiv preprint arXiv:2603.25557}\ } (\bibinfo {year}
  {2026})}\BibitemShut {NoStop}%
\bibitem [{\citenamefont {Giuli}\ \emph {et~al.}(2026)\citenamefont {Giuli},
  \citenamefont {Hasan}, \citenamefont {Kloss}, \citenamefont {Frank},
  \citenamefont {Lee}, \citenamefont {Gingras}, \citenamefont {Yao},\ and\
  \citenamefont {Lanat{\`a}}}]{giuli2026linear}%
  \BibitemOpen
  \bibfield  {author} {\bibinfo {author} {\bibfnamefont {S.}~\bibnamefont
  {Giuli}}, \bibinfo {author} {\bibfnamefont {H.}~\bibnamefont {Hasan}},
  \bibinfo {author} {\bibfnamefont {B.}~\bibnamefont {Kloss}}, \bibinfo
  {author} {\bibfnamefont {M.~S.}\ \bibnamefont {Frank}}, \bibinfo {author}
  {\bibfnamefont {T.-H.}\ \bibnamefont {Lee}}, \bibinfo {author} {\bibfnamefont
  {O.}~\bibnamefont {Gingras}}, \bibinfo {author} {\bibfnamefont {Y.-X.}\
  \bibnamefont {Yao}},\ and\ \bibinfo {author} {\bibfnamefont {N.}~\bibnamefont
  {Lanat{\`a}}},\ }\bibfield  {title} {\bibinfo {title} {Linear foundation
  model for quantum embedding: Data-driven compression of the ghost
  {G}utzwiller variational space},\ }\href {https://doi.org/10.1103/phk3-pvhm}
  {\bibfield  {journal} {\bibinfo  {journal} {PRX Intelligence}\ }\textbf
  {\bibinfo {volume} {1}},\ \bibinfo {pages} {013001} (\bibinfo {year}
  {2026})},\ \bibinfo {note} {arXiv:2512.21666}\BibitemShut {NoStop}%
\bibitem [{\citenamefont {Geiger}\ and\ \citenamefont
  {Smidt}()}]{e3nnGieger2022}%
  \BibitemOpen
  \bibfield  {author} {\bibinfo {author} {\bibfnamefont {M.}~\bibnamefont
  {Geiger}}\ and\ \bibinfo {author} {\bibfnamefont {T.}~\bibnamefont {Smidt}},\
  }\href@noop {} {\bibinfo {title} {e3nn: Euclidean neural networks}},\ \Eprint
  {https://arxiv.org/abs/2207.09453} {arXiv:2207.09453} \BibitemShut {NoStop}%
\bibitem [{\citenamefont {Kondor}\ \emph {et~al.}(2018)\citenamefont {Kondor},
  \citenamefont {Lin},\ and\ \citenamefont {Trivedi}}]{e3nnKondor2018}%
  \BibitemOpen
  \bibfield  {author} {\bibinfo {author} {\bibfnamefont {R.}~\bibnamefont
  {Kondor}}, \bibinfo {author} {\bibfnamefont {Z.}~\bibnamefont {Lin}},\ and\
  \bibinfo {author} {\bibfnamefont {S.}~\bibnamefont {Trivedi}},\ }\bibfield
  {title} {\bibinfo {title} {Clebsch\textendash gordan nets: a fully fourier
  space spherical convolutional neural network},\ }in\ \href
  {https://proceedings.neurips.cc/paper_files/paper/2018/file/a3fc981af450752046be179185ebc8b5-Paper.pdf}
  {\emph {\bibinfo {booktitle} {Adv. Neural Inf. Process. Syst.}}},\
  Vol.~\bibinfo {volume} {31}\ (\bibinfo  {publisher} {Curran Associates,
  Inc.},\ \bibinfo {year} {2018})\BibitemShut {NoStop}%
\bibitem [{\citenamefont {Thomas}\ \emph {et~al.}()\citenamefont {Thomas},
  \citenamefont {Smidt}, \citenamefont {Kearnes}, \citenamefont {Yang},
  \citenamefont {Li}, \citenamefont {Kohlhoff},\ and\ \citenamefont
  {Riley}}]{e3nnThomas2018}%
  \BibitemOpen
  \bibfield  {author} {\bibinfo {author} {\bibfnamefont {N.}~\bibnamefont
  {Thomas}}, \bibinfo {author} {\bibfnamefont {T.}~\bibnamefont {Smidt}},
  \bibinfo {author} {\bibfnamefont {S.}~\bibnamefont {Kearnes}}, \bibinfo
  {author} {\bibfnamefont {L.}~\bibnamefont {Yang}}, \bibinfo {author}
  {\bibfnamefont {L.}~\bibnamefont {Li}}, \bibinfo {author} {\bibfnamefont
  {K.}~\bibnamefont {Kohlhoff}},\ and\ \bibinfo {author} {\bibfnamefont
  {P.}~\bibnamefont {Riley}},\ }\href@noop {} {\bibinfo {title} {Tensor field
  networks: Rotation- and translation-equivariant neural networks for 3d point
  clouds}},\ \Eprint {https://arxiv.org/abs/1802.08219} {arXiv:1802.08219}
  \BibitemShut {NoStop}%
\bibitem [{\citenamefont {Weiler}\ \emph {et~al.}(2018)\citenamefont {Weiler},
  \citenamefont {Geiger}, \citenamefont {Welling}, \citenamefont {Boomsma},\
  and\ \citenamefont {Cohen}}]{e3nnWeiler2018}%
  \BibitemOpen
  \bibfield  {author} {\bibinfo {author} {\bibfnamefont {M.}~\bibnamefont
  {Weiler}}, \bibinfo {author} {\bibfnamefont {M.}~\bibnamefont {Geiger}},
  \bibinfo {author} {\bibfnamefont {M.}~\bibnamefont {Welling}}, \bibinfo
  {author} {\bibfnamefont {W.}~\bibnamefont {Boomsma}},\ and\ \bibinfo {author}
  {\bibfnamefont {T.~S.}\ \bibnamefont {Cohen}},\ }\bibfield  {title} {\bibinfo
  {title} {3\text{D} steerable \text{CNN}s: Learning rotationally equivariant
  features in volumetric data},\ }in\ \href
  {https://proceedings.neurips.cc/paper_files/paper/2018/file/488e4104520c6aab692863cc1dba45af-Paper.pdf}
  {\emph {\bibinfo {booktitle} {Adv. Neural Inf. Process. Syst.}}},\
  Vol.~\bibinfo {volume} {31}\ (\bibinfo  {publisher} {Curran Associates,
  Inc.},\ \bibinfo {year} {2018})\BibitemShut {NoStop}%
\bibitem [{\citenamefont {Boehnke}\ \emph {et~al.}(2011)\citenamefont
  {Boehnke}, \citenamefont {Hafermann}, \citenamefont {Ferrero}, \citenamefont
  {Lechermann},\ and\ \citenamefont {Parcollet}}]{Boehnke2011}%
  \BibitemOpen
  \bibfield  {author} {\bibinfo {author} {\bibfnamefont {L.}~\bibnamefont
  {Boehnke}}, \bibinfo {author} {\bibfnamefont {H.}~\bibnamefont {Hafermann}},
  \bibinfo {author} {\bibfnamefont {M.}~\bibnamefont {Ferrero}}, \bibinfo
  {author} {\bibfnamefont {F.}~\bibnamefont {Lechermann}},\ and\ \bibinfo
  {author} {\bibfnamefont {O.}~\bibnamefont {Parcollet}},\ }\bibfield  {title}
  {\bibinfo {title} {{Orthogonal polynomial representation of imaginary-time
  Green's functions}},\ }\href {https://doi.org/10.1103/PhysRevB.84.075145}
  {\bibfield  {journal} {\bibinfo  {journal} {Phys. Rev. B}\ }\textbf {\bibinfo
  {volume} {84}},\ \bibinfo {pages} {075145} (\bibinfo {year}
  {2011})}\BibitemShut {NoStop}%
\bibitem [{\citenamefont {Wang}\ \emph {et~al.}(2011)\citenamefont {Wang},
  \citenamefont {Dang},\ and\ \citenamefont {Millis}}]{sigmaWang2011}%
  \BibitemOpen
  \bibfield  {author} {\bibinfo {author} {\bibfnamefont {X.}~\bibnamefont
  {Wang}}, \bibinfo {author} {\bibfnamefont {H.~T.}\ \bibnamefont {Dang}},\
  and\ \bibinfo {author} {\bibfnamefont {A.~J.}\ \bibnamefont {Millis}},\
  }\bibfield  {title} {\bibinfo {title} {High-frequency asymptotic behavior of
  self-energies in quantum impurity models},\ }\href
  {https://doi.org/10.1103/PhysRevB.84.073104} {\bibfield  {journal} {\bibinfo
  {journal} {Phys. Rev. B}\ }\textbf {\bibinfo {volume} {84}},\ \bibinfo
  {pages} {073104} (\bibinfo {year} {2011})}\BibitemShut {NoStop}%
\bibitem [{\citenamefont {Zhu}\ \emph {et~al.}(2016)\citenamefont {Zhu},
  \citenamefont {Amsler}, \citenamefont {Fuhrer}, \citenamefont {Schaefer},
  \citenamefont {Faraji}, \citenamefont {Rostami}, \citenamefont {Ghasemi},
  \citenamefont {Sadeghi}, \citenamefont {Grauzinyte}, \citenamefont
  {Wolverton},\ and\ \citenamefont {Goedecker}}]{fpZhu2016}%
  \BibitemOpen
  \bibfield  {author} {\bibinfo {author} {\bibfnamefont {L.}~\bibnamefont
  {Zhu}}, \bibinfo {author} {\bibfnamefont {M.}~\bibnamefont {Amsler}},
  \bibinfo {author} {\bibfnamefont {T.}~\bibnamefont {Fuhrer}}, \bibinfo
  {author} {\bibfnamefont {B.}~\bibnamefont {Schaefer}}, \bibinfo {author}
  {\bibfnamefont {S.}~\bibnamefont {Faraji}}, \bibinfo {author} {\bibfnamefont
  {S.}~\bibnamefont {Rostami}}, \bibinfo {author} {\bibfnamefont {S.~A.}\
  \bibnamefont {Ghasemi}}, \bibinfo {author} {\bibfnamefont {A.}~\bibnamefont
  {Sadeghi}}, \bibinfo {author} {\bibfnamefont {M.}~\bibnamefont {Grauzinyte}},
  \bibinfo {author} {\bibfnamefont {C.}~\bibnamefont {Wolverton}},\ and\
  \bibinfo {author} {\bibfnamefont {S.}~\bibnamefont {Goedecker}},\ }\bibfield
  {title} {\bibinfo {title} {A fingerprint based metric for measuring
  similarities of crystalline structures},\ }\href
  {https://doi.org/10.1063/1.4940026} {\bibfield  {journal} {\bibinfo
  {journal} {J. Chem. Phys.}\ }\textbf {\bibinfo {volume} {144}},\ \bibinfo
  {pages} {034203} (\bibinfo {year} {2016})}\BibitemShut {NoStop}%
\bibitem [{\citenamefont {Anzellini}\ \emph {et~al.}(2013)\citenamefont
  {Anzellini}, \citenamefont {Dewaele}, \citenamefont {Mezouar}, \citenamefont
  {Loubeyre},\ and\ \citenamefont {Morard}}]{emcAnzellini2013}%
  \BibitemOpen
  \bibfield  {author} {\bibinfo {author} {\bibfnamefont {S.}~\bibnamefont
  {Anzellini}}, \bibinfo {author} {\bibfnamefont {A.}~\bibnamefont {Dewaele}},
  \bibinfo {author} {\bibfnamefont {M.}~\bibnamefont {Mezouar}}, \bibinfo
  {author} {\bibfnamefont {P.}~\bibnamefont {Loubeyre}},\ and\ \bibinfo
  {author} {\bibfnamefont {G.}~\bibnamefont {Morard}},\ }\bibfield  {title}
  {\bibinfo {title} {Melting of iron at earth’s inner core boundary based on
  fast x-ray diffraction},\ }\href {https://doi.org/10.1126/science.1233514}
  {\bibfield  {journal} {\bibinfo  {journal} {Science}\ }\textbf {\bibinfo
  {volume} {340}},\ \bibinfo {pages} {464} (\bibinfo {year}
  {2013})}\BibitemShut {NoStop}%
\bibitem [{\citenamefont {Kraus}\ \emph {et~al.}(2022)\citenamefont {Kraus},
  \citenamefont {Hemley}, \citenamefont {Ali}, \citenamefont {Belof},
  \citenamefont {Benedict}, \citenamefont {Bernier}, \citenamefont {Braun},
  \citenamefont {Cohen}, \citenamefont {Collins}, \citenamefont {Coppari},
  \citenamefont {Desjarlais}, \citenamefont {Fratanduono}, \citenamefont
  {Hamel}, \citenamefont {Krygier}, \citenamefont {Lazicki}, \citenamefont
  {Mcnaney}, \citenamefont {Millot}, \citenamefont {Myint}, \citenamefont
  {Newman}, \citenamefont {Rygg}, \citenamefont {Sterbentz}, \citenamefont
  {Stewart}, \citenamefont {Stixrude}, \citenamefont {Swift}, \citenamefont
  {Wehrenberg},\ and\ \citenamefont {Eggert}}]{emcKraus2022}%
  \BibitemOpen
  \bibfield  {author} {\bibinfo {author} {\bibfnamefont {R.~G.}\ \bibnamefont
  {Kraus}}, \bibinfo {author} {\bibfnamefont {R.~J.}\ \bibnamefont {Hemley}},
  \bibinfo {author} {\bibfnamefont {S.~J.}\ \bibnamefont {Ali}}, \bibinfo
  {author} {\bibfnamefont {J.~L.}\ \bibnamefont {Belof}}, \bibinfo {author}
  {\bibfnamefont {L.~X.}\ \bibnamefont {Benedict}}, \bibinfo {author}
  {\bibfnamefont {J.}~\bibnamefont {Bernier}}, \bibinfo {author} {\bibfnamefont
  {D.}~\bibnamefont {Braun}}, \bibinfo {author} {\bibfnamefont {R.~E.}\
  \bibnamefont {Cohen}}, \bibinfo {author} {\bibfnamefont {G.~W.}\ \bibnamefont
  {Collins}}, \bibinfo {author} {\bibfnamefont {F.}~\bibnamefont {Coppari}},
  \bibinfo {author} {\bibfnamefont {M.~P.}\ \bibnamefont {Desjarlais}},
  \bibinfo {author} {\bibfnamefont {D.}~\bibnamefont {Fratanduono}}, \bibinfo
  {author} {\bibfnamefont {S.}~\bibnamefont {Hamel}}, \bibinfo {author}
  {\bibfnamefont {A.}~\bibnamefont {Krygier}}, \bibinfo {author} {\bibfnamefont
  {A.}~\bibnamefont {Lazicki}}, \bibinfo {author} {\bibfnamefont
  {J.}~\bibnamefont {Mcnaney}}, \bibinfo {author} {\bibfnamefont
  {M.}~\bibnamefont {Millot}}, \bibinfo {author} {\bibfnamefont {P.~C.}\
  \bibnamefont {Myint}}, \bibinfo {author} {\bibfnamefont {M.~G.}\ \bibnamefont
  {Newman}}, \bibinfo {author} {\bibfnamefont {J.~R.}\ \bibnamefont {Rygg}},
  \bibinfo {author} {\bibfnamefont {D.~M.}\ \bibnamefont {Sterbentz}}, \bibinfo
  {author} {\bibfnamefont {S.~T.}\ \bibnamefont {Stewart}}, \bibinfo {author}
  {\bibfnamefont {L.}~\bibnamefont {Stixrude}}, \bibinfo {author}
  {\bibfnamefont {D.~C.}\ \bibnamefont {Swift}}, \bibinfo {author}
  {\bibfnamefont {C.}~\bibnamefont {Wehrenberg}},\ and\ \bibinfo {author}
  {\bibfnamefont {J.~H.}\ \bibnamefont {Eggert}},\ }\bibfield  {title}
  {\bibinfo {title} {Measuring the melting curve of iron at super-earth core
  conditions},\ }\href {https://doi.org/10.1126/science.abm1472} {\bibfield
  {journal} {\bibinfo  {journal} {Science}\ }\textbf {\bibinfo {volume}
  {375}},\ \bibinfo {pages} {202} (\bibinfo {year} {2022})}\BibitemShut
  {NoStop}%
\bibitem [{\citenamefont {Balugani}\ \emph {et~al.}(2024)\citenamefont
  {Balugani}, \citenamefont {Hernandez}, \citenamefont {S\'evelin-Radiguet},
  \citenamefont {Mathon}, \citenamefont {Recoules}, \citenamefont {Kas},
  \citenamefont {Eakins}, \citenamefont {Doyle}, \citenamefont {Ravasio},\ and\
  \citenamefont {Torchio}}]{emcBalugani2024}%
  \BibitemOpen
  \bibfield  {author} {\bibinfo {author} {\bibfnamefont {S.}~\bibnamefont
  {Balugani}}, \bibinfo {author} {\bibfnamefont {J.~A.}\ \bibnamefont
  {Hernandez}}, \bibinfo {author} {\bibfnamefont {N.}~\bibnamefont
  {S\'evelin-Radiguet}}, \bibinfo {author} {\bibfnamefont {O.}~\bibnamefont
  {Mathon}}, \bibinfo {author} {\bibfnamefont {V.}~\bibnamefont {Recoules}},
  \bibinfo {author} {\bibfnamefont {J.~J.}\ \bibnamefont {Kas}}, \bibinfo
  {author} {\bibfnamefont {D.~E.}\ \bibnamefont {Eakins}}, \bibinfo {author}
  {\bibfnamefont {H.}~\bibnamefont {Doyle}}, \bibinfo {author} {\bibfnamefont
  {A.}~\bibnamefont {Ravasio}},\ and\ \bibinfo {author} {\bibfnamefont
  {R.}~\bibnamefont {Torchio}},\ }\bibfield  {title} {\bibinfo {title} {New
  constraints on the melting temperature and phase stability of shocked iron up
  to 270 gpa probed by ultrafast x-ray absorption spectroscopy},\ }\href
  {https://doi.org/10.1103/PhysRevLett.133.254101} {\bibfield  {journal}
  {\bibinfo  {journal} {Phys. Rev. Lett.}\ }\textbf {\bibinfo {volume} {133}},\
  \bibinfo {pages} {254101} (\bibinfo {year} {2024})}\BibitemShut {NoStop}%
\bibitem [{\citenamefont {Sinmyo}\ \emph {et~al.}(2019)\citenamefont {Sinmyo},
  \citenamefont {Hirose},\ and\ \citenamefont {Ohishi}}]{emcSinmyo2019}%
  \BibitemOpen
  \bibfield  {author} {\bibinfo {author} {\bibfnamefont {R.}~\bibnamefont
  {Sinmyo}}, \bibinfo {author} {\bibfnamefont {K.}~\bibnamefont {Hirose}},\
  and\ \bibinfo {author} {\bibfnamefont {Y.}~\bibnamefont {Ohishi}},\
  }\bibfield  {title} {\bibinfo {title} {Melting curve of iron to 290
  \text{GPa} determined in a resistance-heated diamond-anvil cell},\ }\href
  {https://doi.org/https://doi.org/10.1016/j.epsl.2019.01.006} {\bibfield
  {journal} {\bibinfo  {journal} {Earth Planet. Sci. Lett.}\ }\textbf {\bibinfo
  {volume} {510}},\ \bibinfo {pages} {45} (\bibinfo {year} {2019})}\BibitemShut
  {NoStop}%
\bibitem [{\citenamefont {Alf\`e}\ \emph {et~al.}(2002)\citenamefont {Alf\`e},
  \citenamefont {Price},\ and\ \citenamefont {Gillan}}]{aimcAlfe2002}%
  \BibitemOpen
  \bibfield  {author} {\bibinfo {author} {\bibfnamefont {D.}~\bibnamefont
  {Alf\`e}}, \bibinfo {author} {\bibfnamefont {G.~D.}\ \bibnamefont {Price}},\
  and\ \bibinfo {author} {\bibfnamefont {M.~J.}\ \bibnamefont {Gillan}},\
  }\bibfield  {title} {\bibinfo {title} {Iron under earth's core conditions:
  Liquid-state thermodynamics and high-pressure melting curve from ab initio
  calculations},\ }\href {https://doi.org/10.1103/PhysRevB.65.165118}
  {\bibfield  {journal} {\bibinfo  {journal} {Phys. Rev. B}\ }\textbf {\bibinfo
  {volume} {65}},\ \bibinfo {pages} {165118} (\bibinfo {year}
  {2002})}\BibitemShut {NoStop}%
\bibitem [{\citenamefont {Alf\`e}(2009)}]{aimcAlfe2009}%
  \BibitemOpen
  \bibfield  {author} {\bibinfo {author} {\bibfnamefont {D.}~\bibnamefont
  {Alf\`e}},\ }\bibfield  {title} {\bibinfo {title} {Temperature of the
  inner-core boundary of the earth: Melting of iron at high pressure from
  first-principles coexistence simulations},\ }\href
  {https://doi.org/10.1103/PhysRevB.79.060101} {\bibfield  {journal} {\bibinfo
  {journal} {Phys. Rev. B}\ }\textbf {\bibinfo {volume} {79}},\ \bibinfo
  {pages} {060101} (\bibinfo {year} {2009})}\BibitemShut {NoStop}%
\bibitem [{\citenamefont {Belonoshko}\ \emph {et~al.}(2021)\citenamefont
  {Belonoshko}, \citenamefont {Fu},\ and\ \citenamefont
  {Smirnov}}]{aimcBelonoshko2021}%
  \BibitemOpen
  \bibfield  {author} {\bibinfo {author} {\bibfnamefont {A.~B.}\ \bibnamefont
  {Belonoshko}}, \bibinfo {author} {\bibfnamefont {J.}~\bibnamefont {Fu}},\
  and\ \bibinfo {author} {\bibfnamefont {G.}~\bibnamefont {Smirnov}},\
  }\bibfield  {title} {\bibinfo {title} {Free energies of iron phases at high
  pressure and temperature: Molecular dynamics study},\ }\href
  {https://doi.org/10.1103/PhysRevB.104.104103} {\bibfield  {journal} {\bibinfo
   {journal} {Phys. Rev. B}\ }\textbf {\bibinfo {volume} {104}},\ \bibinfo
  {pages} {104103} (\bibinfo {year} {2021})}\BibitemShut {NoStop}%
\bibitem [{\citenamefont {Belonoshko}\ \emph {et~al.}(2000)\citenamefont
  {Belonoshko}, \citenamefont {Ahuja},\ and\ \citenamefont
  {Johansson}}]{aimcBelonosko2000}%
  \BibitemOpen
  \bibfield  {author} {\bibinfo {author} {\bibfnamefont {A.~B.}\ \bibnamefont
  {Belonoshko}}, \bibinfo {author} {\bibfnamefont {R.}~\bibnamefont {Ahuja}},\
  and\ \bibinfo {author} {\bibfnamefont {B.}~\bibnamefont {Johansson}},\
  }\bibfield  {title} {\bibinfo {title} {Quasi--ab initio molecular dynamic
  study of \text{Fe} melting},\ }\href
  {https://doi.org/10.1103/PhysRevLett.84.3638} {\bibfield  {journal} {\bibinfo
   {journal} {Phys. Rev. Lett.}\ }\textbf {\bibinfo {volume} {84}},\ \bibinfo
  {pages} {3638} (\bibinfo {year} {2000})}\BibitemShut {NoStop}%
\bibitem [{\citenamefont {Gonz\'alez-Cataldo}\ and\ \citenamefont
  {Militzer}(2023)}]{aimcGonzalez2023}%
  \BibitemOpen
  \bibfield  {author} {\bibinfo {author} {\bibfnamefont {F.}~\bibnamefont
  {Gonz\'alez-Cataldo}}\ and\ \bibinfo {author} {\bibfnamefont
  {B.}~\bibnamefont {Militzer}},\ }\bibfield  {title} {\bibinfo {title} {Ab
  initio determination of iron melting at terapascal pressures and super-earths
  core crystallization},\ }\href
  {https://doi.org/10.1103/PhysRevResearch.5.033194} {\bibfield  {journal}
  {\bibinfo  {journal} {Phys. Rev. Res.}\ }\textbf {\bibinfo {volume} {5}},\
  \bibinfo {pages} {033194} (\bibinfo {year} {2023})}\BibitemShut {NoStop}%
\bibitem [{\citenamefont {Sola}\ and\ \citenamefont
  {Alf\`e}(2009)}]{aimcSola2009}%
  \BibitemOpen
  \bibfield  {author} {\bibinfo {author} {\bibfnamefont {E.}~\bibnamefont
  {Sola}}\ and\ \bibinfo {author} {\bibfnamefont {D.}~\bibnamefont {Alf\`e}},\
  }\bibfield  {title} {\bibinfo {title} {Melting of iron under earth's core
  conditions from diffusion monte carlo free energy calculations},\ }\href
  {https://doi.org/10.1103/PhysRevLett.103.078501} {\bibfield  {journal}
  {\bibinfo  {journal} {Phys. Rev. Lett.}\ }\textbf {\bibinfo {volume} {103}},\
  \bibinfo {pages} {078501} (\bibinfo {year} {2009})}\BibitemShut {NoStop}%
\bibitem [{\citenamefont {Stixrude}(2014)}]{aimcStixrude2014}%
  \BibitemOpen
  \bibfield  {author} {\bibinfo {author} {\bibfnamefont {L.}~\bibnamefont
  {Stixrude}},\ }\bibfield  {title} {\bibinfo {title} {Melting in
  super-earths},\ }\href {https://doi.org/10.1098/rsta.2013.0076} {\bibfield
  {journal} {\bibinfo  {journal} {Philos. Trans. R. Soc. A}\ }\textbf {\bibinfo
  {volume} {372}},\ \bibinfo {pages} {20130076} (\bibinfo {year}
  {2014})}\BibitemShut {NoStop}%
\bibitem [{\citenamefont {Sun}\ \emph {et~al.}(2022)\citenamefont {Sun},
  \citenamefont {Zhang}, \citenamefont {Mendelev}, \citenamefont
  {Wentzcovitch},\ and\ \citenamefont {Ho}}]{aimcSun2022}%
  \BibitemOpen
  \bibfield  {author} {\bibinfo {author} {\bibfnamefont {Y.}~\bibnamefont
  {Sun}}, \bibinfo {author} {\bibfnamefont {F.}~\bibnamefont {Zhang}}, \bibinfo
  {author} {\bibfnamefont {M.~I.}\ \bibnamefont {Mendelev}}, \bibinfo {author}
  {\bibfnamefont {R.~M.}\ \bibnamefont {Wentzcovitch}},\ and\ \bibinfo {author}
  {\bibfnamefont {K.-M.}\ \bibnamefont {Ho}},\ }\bibfield  {title} {\bibinfo
  {title} {Two-step nucleation of the earth’s inner core},\ }\href
  {https://doi.org/10.1073/pnas.2113059119} {\bibfield  {journal} {\bibinfo
  {journal} {Proc. Natl. Acad. Sci. U.S.A.}\ }\textbf {\bibinfo {volume}
  {119}},\ \bibinfo {pages} {e2113059119} (\bibinfo {year} {2022})}\BibitemShut
  {NoStop}%
\bibitem [{\citenamefont {Sun}\ \emph {et~al.}(2023)\citenamefont {Sun},
  \citenamefont {Mendelev}, \citenamefont {Zhang}, \citenamefont {Liu},
  \citenamefont {Da}, \citenamefont {Wang}, \citenamefont {Wentzcovitch},\ and\
  \citenamefont {Ho}}]{aimcSun2023}%
  \BibitemOpen
  \bibfield  {author} {\bibinfo {author} {\bibfnamefont {Y.}~\bibnamefont
  {Sun}}, \bibinfo {author} {\bibfnamefont {M.~I.}\ \bibnamefont {Mendelev}},
  \bibinfo {author} {\bibfnamefont {F.}~\bibnamefont {Zhang}}, \bibinfo
  {author} {\bibfnamefont {X.}~\bibnamefont {Liu}}, \bibinfo {author}
  {\bibfnamefont {B.}~\bibnamefont {Da}}, \bibinfo {author} {\bibfnamefont
  {C.-Z.}\ \bibnamefont {Wang}}, \bibinfo {author} {\bibfnamefont {R.~M.}\
  \bibnamefont {Wentzcovitch}},\ and\ \bibinfo {author} {\bibfnamefont {K.-M.}\
  \bibnamefont {Ho}},\ }\bibfield  {title} {\bibinfo {title} {Ab initio melting
  temperatures of bcc and hcp iron under the earth's inner core condition},\
  }\href {https://doi.org/https://doi.org/10.1029/2022GL102447} {\bibfield
  {journal} {\bibinfo  {journal} {Geophys. Res. Lett.}\ }\textbf {\bibinfo
  {volume} {50}},\ \bibinfo {pages} {e2022GL102447} (\bibinfo {year}
  {2023})}\BibitemShut {NoStop}%
\bibitem [{\citenamefont {Wu}\ \emph {et~al.}(2024)\citenamefont {Wu},
  \citenamefont {Wu}, \citenamefont {Wang}, \citenamefont {Ho}, \citenamefont
  {Wentzcovitch},\ and\ \citenamefont {Sun}}]{aimcWu2024}%
  \BibitemOpen
  \bibfield  {author} {\bibinfo {author} {\bibfnamefont {F.}~\bibnamefont
  {Wu}}, \bibinfo {author} {\bibfnamefont {S.}~\bibnamefont {Wu}}, \bibinfo
  {author} {\bibfnamefont {C.-Z.}\ \bibnamefont {Wang}}, \bibinfo {author}
  {\bibfnamefont {K.-M.}\ \bibnamefont {Ho}}, \bibinfo {author} {\bibfnamefont
  {R.~M.}\ \bibnamefont {Wentzcovitch}},\ and\ \bibinfo {author} {\bibfnamefont
  {Y.}~\bibnamefont {Sun}},\ }\bibfield  {title} {\bibinfo {title} {Melting
  temperature of iron under the earth’s inner core condition from deep
  machine learning},\ }\href
  {https://doi.org/https://doi.org/10.1016/j.gsf.2024.101925} {\bibfield
  {journal} {\bibinfo  {journal} {Geosci. Front.}\ }\textbf {\bibinfo {volume}
  {15}},\ \bibinfo {pages} {101925} (\bibinfo {year} {2024})}\BibitemShut
  {NoStop}%
\bibitem [{\citenamefont {Sun}\ \emph {et~al.}(2018)\citenamefont {Sun},
  \citenamefont {Brodholt}, \citenamefont {Li},\ and\ \citenamefont {Vo{\v
  c}adlo}}]{fepSun2018}%
  \BibitemOpen
  \bibfield  {author} {\bibinfo {author} {\bibfnamefont {T.}~\bibnamefont
  {Sun}}, \bibinfo {author} {\bibfnamefont {J.~P.}\ \bibnamefont {Brodholt}},
  \bibinfo {author} {\bibfnamefont {Y.}~\bibnamefont {Li}},\ and\ \bibinfo
  {author} {\bibfnamefont {L.}~\bibnamefont {Vo{\v c}adlo}},\ }\bibfield
  {title} {\bibinfo {title} {Melting properties from ab initio free energy
  calculations: Iron at the {Earth}'s inner-core boundary},\ }\href
  {https://doi.org/10.1103/PhysRevB.98.224301} {\bibfield  {journal} {\bibinfo
  {journal} {Phys. Rev. B}\ }\textbf {\bibinfo {volume} {98}},\ \bibinfo
  {pages} {224301} (\bibinfo {year} {2018})}\BibitemShut {NoStop}%
\bibitem [{\citenamefont {Pourovskii}\ \emph {et~al.}(2017)\citenamefont
  {Pourovskii}, \citenamefont {Mravlje}, \citenamefont {Georges}, \citenamefont
  {Simak},\ and\ \citenamefont {Abrikosov}}]{ieccPourovskii2017}%
  \BibitemOpen
  \bibfield  {author} {\bibinfo {author} {\bibfnamefont {L.~V.}\ \bibnamefont
  {Pourovskii}}, \bibinfo {author} {\bibfnamefont {J.}~\bibnamefont {Mravlje}},
  \bibinfo {author} {\bibfnamefont {A.}~\bibnamefont {Georges}}, \bibinfo
  {author} {\bibfnamefont {S.~I.}\ \bibnamefont {Simak}},\ and\ \bibinfo
  {author} {\bibfnamefont {I.~A.}\ \bibnamefont {Abrikosov}},\ }\bibfield
  {title} {\bibinfo {title} {Electron–electron scattering and thermal
  conductivity of $\epsilon$-iron at earth’s core conditions},\ }\href
  {https://doi.org/10.1088/1367-2630/aa76c9} {\bibfield  {journal} {\bibinfo
  {journal} {New J. Phys.}\ }\textbf {\bibinfo {volume} {19}},\ \bibinfo
  {pages} {073022} (\bibinfo {year} {2017})}\BibitemShut {NoStop}%
\bibitem [{\citenamefont {Pourovskii}(2019)}]{ieccPourovskii2019}%
  \BibitemOpen
  \bibfield  {author} {\bibinfo {author} {\bibfnamefont {L.~V.}\ \bibnamefont
  {Pourovskii}},\ }\bibfield  {title} {\bibinfo {title} {Electronic
  correlations in dense iron: from moderate pressure to earth’s core
  conditions},\ }\href {https://doi.org/10.1088/1361-648X/ab274f} {\bibfield
  {journal} {\bibinfo  {journal} {J. Phys. Condens. Matter}\ }\textbf {\bibinfo
  {volume} {31}},\ \bibinfo {pages} {373001} (\bibinfo {year}
  {2019})}\BibitemShut {NoStop}%
\bibitem [{\citenamefont {Vekilova}\ \emph {et~al.}(2015)\citenamefont
  {Vekilova}, \citenamefont {Pourovskii}, \citenamefont {Abrikosov},\ and\
  \citenamefont {Simak}}]{ieccVekilova2015}%
  \BibitemOpen
  \bibfield  {author} {\bibinfo {author} {\bibfnamefont {O.~Y.}\ \bibnamefont
  {Vekilova}}, \bibinfo {author} {\bibfnamefont {L.~V.}\ \bibnamefont
  {Pourovskii}}, \bibinfo {author} {\bibfnamefont {I.~A.}\ \bibnamefont
  {Abrikosov}},\ and\ \bibinfo {author} {\bibfnamefont {S.~I.}\ \bibnamefont
  {Simak}},\ }\bibfield  {title} {\bibinfo {title} {Electronic correlations in
  \text{Fe} at earth's inner core conditions: Effects of alloying with
  \text{Ni}},\ }\href {https://doi.org/10.1103/PhysRevB.91.245116} {\bibfield
  {journal} {\bibinfo  {journal} {Phys. Rev. B}\ }\textbf {\bibinfo {volume}
  {91}},\ \bibinfo {pages} {245116} (\bibinfo {year} {2015})}\BibitemShut
  {NoStop}%
\bibitem [{\citenamefont {Tan}\ \emph {et~al.}()\citenamefont {Tan},
  \citenamefont {Descoteaux}, \citenamefont {Kotak}, \citenamefont
  {de~Miranda~Nascimento}, \citenamefont {Kavanagh}, \citenamefont {Zichi},
  \citenamefont {Wang}, \citenamefont {Saluja}, \citenamefont {Hu},
  \citenamefont {Smidt}, \citenamefont {Johansson}, \citenamefont {Witt},
  \citenamefont {Kozinsky},\ and\ \citenamefont {Musaelian}}]{nequipTan2025}%
  \BibitemOpen
  \bibfield  {author} {\bibinfo {author} {\bibfnamefont {C.~W.}\ \bibnamefont
  {Tan}}, \bibinfo {author} {\bibfnamefont {M.~L.}\ \bibnamefont {Descoteaux}},
  \bibinfo {author} {\bibfnamefont {M.}~\bibnamefont {Kotak}}, \bibinfo
  {author} {\bibfnamefont {G.}~\bibnamefont {de~Miranda~Nascimento}}, \bibinfo
  {author} {\bibfnamefont {S.~R.}\ \bibnamefont {Kavanagh}}, \bibinfo {author}
  {\bibfnamefont {L.}~\bibnamefont {Zichi}}, \bibinfo {author} {\bibfnamefont
  {M.}~\bibnamefont {Wang}}, \bibinfo {author} {\bibfnamefont {A.}~\bibnamefont
  {Saluja}}, \bibinfo {author} {\bibfnamefont {Y.~R.}\ \bibnamefont {Hu}},
  \bibinfo {author} {\bibfnamefont {T.}~\bibnamefont {Smidt}}, \bibinfo
  {author} {\bibfnamefont {A.}~\bibnamefont {Johansson}}, \bibinfo {author}
  {\bibfnamefont {W.~C.}\ \bibnamefont {Witt}}, \bibinfo {author}
  {\bibfnamefont {B.}~\bibnamefont {Kozinsky}},\ and\ \bibinfo {author}
  {\bibfnamefont {A.}~\bibnamefont {Musaelian}},\ }\href@noop {} {\bibinfo
  {title} {High-performance training and inference for deep equivariant
  interatomic potentials}},\ \Eprint {https://arxiv.org/abs/2504.16068}
  {arXiv:2504.16068} \BibitemShut {NoStop}%
\bibitem [{\citenamefont {Batzner}\ \emph {et~al.}(2022)\citenamefont
  {Batzner}, \citenamefont {Musaelian}, \citenamefont {Sun}, \citenamefont
  {Geiger}, \citenamefont {Mailoa}, \citenamefont {Kornbluth}, \citenamefont
  {Molinari}, \citenamefont {Smidt},\ and\ \citenamefont
  {Kozinsky}}]{nequipBatzner2022}%
  \BibitemOpen
  \bibfield  {author} {\bibinfo {author} {\bibfnamefont {S.}~\bibnamefont
  {Batzner}}, \bibinfo {author} {\bibfnamefont {A.}~\bibnamefont {Musaelian}},
  \bibinfo {author} {\bibfnamefont {L.}~\bibnamefont {Sun}}, \bibinfo {author}
  {\bibfnamefont {M.}~\bibnamefont {Geiger}}, \bibinfo {author} {\bibfnamefont
  {J.~P.}\ \bibnamefont {Mailoa}}, \bibinfo {author} {\bibfnamefont
  {M.}~\bibnamefont {Kornbluth}}, \bibinfo {author} {\bibfnamefont
  {N.}~\bibnamefont {Molinari}}, \bibinfo {author} {\bibfnamefont {T.~E.}\
  \bibnamefont {Smidt}},\ and\ \bibinfo {author} {\bibfnamefont
  {B.}~\bibnamefont {Kozinsky}},\ }\bibfield  {title} {\bibinfo {title}
  {E(3)-equivariant graph neural networks for data-efficient and accurate
  interatomic potentials},\ }\href {https://doi.org/10.1038/s41467-022-29939-5}
  {\bibfield  {journal} {\bibinfo  {journal} {Nat. Commun.}\ }\textbf {\bibinfo
  {volume} {13}},\ \bibinfo {pages} {2453} (\bibinfo {year}
  {2022})}\BibitemShut {NoStop}%
\bibitem [{\citenamefont {Huang}\ and\ \citenamefont
  {Rubenstein}(2023)}]{huangMachineLearningDiffusion2023}%
  \BibitemOpen
  \bibfield  {author} {\bibinfo {author} {\bibfnamefont {C.}~\bibnamefont
  {Huang}}\ and\ \bibinfo {author} {\bibfnamefont {B.~M.}\ \bibnamefont
  {Rubenstein}},\ }\bibfield  {title} {\bibinfo {title} {Machine {{Learning
  Diffusion Monte Carlo Forces}}},\ }\href
  {https://doi.org/10.1021/acs.jpca.2c05904} {\bibfield  {journal} {\bibinfo
  {journal} {J. Phys. Chem. A}\ }\textbf {\bibinfo {volume} {127}},\ \bibinfo
  {pages} {339} (\bibinfo {year} {2023})}\BibitemShut {NoStop}%
\bibitem [{\citenamefont {Deng}\ and\ \citenamefont
  {Stixrude}(2021)}]{pmpDeng2021}%
  \BibitemOpen
  \bibfield  {author} {\bibinfo {author} {\bibfnamefont {J.}~\bibnamefont
  {Deng}}\ and\ \bibinfo {author} {\bibfnamefont {L.}~\bibnamefont
  {Stixrude}},\ }\bibfield  {title} {\bibinfo {title} {Thermal conductivity of
  silicate liquid determined by machine learning potentials},\ }\href
  {https://doi.org/https://doi.org/10.1029/2021GL093806} {\bibfield  {journal}
  {\bibinfo  {journal} {Geophysical Research Letters}\ }\textbf {\bibinfo
  {volume} {48}},\ \bibinfo {pages} {e2021GL093806} (\bibinfo {year} {2021})},\
  \bibinfo {note} {e2021GL093806 2021GL093806}\BibitemShut {NoStop}%
\bibitem [{\citenamefont {Deng}\ \emph {et~al.}(2023)\citenamefont {Deng},
  \citenamefont {Niu}, \citenamefont {Hu}, \citenamefont {Chen},\ and\
  \citenamefont {Stixrude}}]{pmpDeng2023}%
  \BibitemOpen
  \bibfield  {author} {\bibinfo {author} {\bibfnamefont {J.}~\bibnamefont
  {Deng}}, \bibinfo {author} {\bibfnamefont {H.}~\bibnamefont {Niu}}, \bibinfo
  {author} {\bibfnamefont {J.}~\bibnamefont {Hu}}, \bibinfo {author}
  {\bibfnamefont {M.}~\bibnamefont {Chen}},\ and\ \bibinfo {author}
  {\bibfnamefont {L.}~\bibnamefont {Stixrude}},\ }\bibfield  {title} {\bibinfo
  {title} {Melting of {$\mathrm{MgSi}{\mathrm{O}}_{3}$} determined by machine
  learning potentials},\ }\href {https://doi.org/10.1103/PhysRevB.107.064103}
  {\bibfield  {journal} {\bibinfo  {journal} {Phys. Rev. B}\ }\textbf {\bibinfo
  {volume} {107}},\ \bibinfo {pages} {064103} (\bibinfo {year}
  {2023})}\BibitemShut {NoStop}%
\bibitem [{\citenamefont {Fan}\ \emph {et~al.}(2025)\citenamefont {Fan},
  \citenamefont {Whittaker},\ and\ \citenamefont {Asta}}]{pmpFan2025}%
  \BibitemOpen
  \bibfield  {author} {\bibinfo {author} {\bibfnamefont {Z.}~\bibnamefont
  {Fan}}, \bibinfo {author} {\bibfnamefont {M.~L.}\ \bibnamefont {Whittaker}},\
  and\ \bibinfo {author} {\bibfnamefont {M.}~\bibnamefont {Asta}},\ }\bibfield
  {title} {\bibinfo {title} {Efficient machine learning interatomic potentials
  robust for liquid and multiple solid polymorphs of {NaF} and {KF}},\ }\href
  {https://doi.org/10.1103/xbfm-clgd} {\bibfield  {journal} {\bibinfo
  {journal} {Phys. Rev. Mater.}\ }\textbf {\bibinfo {volume} {9}},\ \bibinfo
  {pages} {103406} (\bibinfo {year} {2025})}\BibitemShut {NoStop}%
\bibitem [{\citenamefont {Pun}\ \emph {et~al.}(2020)\citenamefont {Pun},
  \citenamefont {Yamakov}, \citenamefont {Hickman}, \citenamefont {Glaessgen},\
  and\ \citenamefont {Mishin}}]{pmpPun2020}%
  \BibitemOpen
  \bibfield  {author} {\bibinfo {author} {\bibfnamefont {G.~P.~P.}\
  \bibnamefont {Pun}}, \bibinfo {author} {\bibfnamefont {V.}~\bibnamefont
  {Yamakov}}, \bibinfo {author} {\bibfnamefont {J.}~\bibnamefont {Hickman}},
  \bibinfo {author} {\bibfnamefont {E.~H.}\ \bibnamefont {Glaessgen}},\ and\
  \bibinfo {author} {\bibfnamefont {Y.}~\bibnamefont {Mishin}},\ }\bibfield
  {title} {\bibinfo {title} {Development of a general-purpose machine-learning
  interatomic potential for aluminum by the physically informed neural network
  method},\ }\href {https://doi.org/10.1103/PhysRevMaterials.4.113807}
  {\bibfield  {journal} {\bibinfo  {journal} {Phys. Rev. Mater.}\ }\textbf
  {\bibinfo {volume} {4}},\ \bibinfo {pages} {113807} (\bibinfo {year}
  {2020})}\BibitemShut {NoStop}%
\bibitem [{\citenamefont {Tang}\ \emph {et~al.}(2025)\citenamefont {Tang},
  \citenamefont {Xia}, \citenamefont {Viswanathan}, \citenamefont {Soto},
  \citenamefont {Kovnir},\ and\ \citenamefont {Wang}}]{pmpTang2025}%
  \BibitemOpen
  \bibfield  {author} {\bibinfo {author} {\bibfnamefont {L.}~\bibnamefont
  {Tang}}, \bibinfo {author} {\bibfnamefont {W.}~\bibnamefont {Xia}}, \bibinfo
  {author} {\bibfnamefont {G.}~\bibnamefont {Viswanathan}}, \bibinfo {author}
  {\bibfnamefont {E.}~\bibnamefont {Soto}}, \bibinfo {author} {\bibfnamefont
  {K.}~\bibnamefont {Kovnir}},\ and\ \bibinfo {author} {\bibfnamefont {C.-Z.}\
  \bibnamefont {Wang}},\ }\bibfield  {title} {\bibinfo {title} {Developing a
  neural network machine learning interatomic potential for molecular dynamics
  simulations of {La–Si–P} systems},\ }\href
  {https://doi.org/10.1063/5.0284672} {\bibfield  {journal} {\bibinfo
  {journal} {The Journal of Chemical Physics}\ }\textbf {\bibinfo {volume}
  {163}},\ \bibinfo {pages} {084109} (\bibinfo {year} {2025})}\BibitemShut
  {NoStop}%
\bibitem [{\citenamefont {Willman}\ \emph {et~al.}(2022)\citenamefont
  {Willman}, \citenamefont {Nguyen-Cong}, \citenamefont {Williams},
  \citenamefont {Belonoshko}, \citenamefont {Moore}, \citenamefont {Thompson},
  \citenamefont {Wood},\ and\ \citenamefont {Oleynik}}]{pmpWillman2022}%
  \BibitemOpen
  \bibfield  {author} {\bibinfo {author} {\bibfnamefont {J.~T.}\ \bibnamefont
  {Willman}}, \bibinfo {author} {\bibfnamefont {K.}~\bibnamefont
  {Nguyen-Cong}}, \bibinfo {author} {\bibfnamefont {A.~S.}\ \bibnamefont
  {Williams}}, \bibinfo {author} {\bibfnamefont {A.~B.}\ \bibnamefont
  {Belonoshko}}, \bibinfo {author} {\bibfnamefont {S.~G.}\ \bibnamefont
  {Moore}}, \bibinfo {author} {\bibfnamefont {A.~P.}\ \bibnamefont {Thompson}},
  \bibinfo {author} {\bibfnamefont {M.~A.}\ \bibnamefont {Wood}},\ and\
  \bibinfo {author} {\bibfnamefont {I.~I.}\ \bibnamefont {Oleynik}},\
  }\bibfield  {title} {\bibinfo {title} {Machine learning interatomic potential
  for simulations of carbon at extreme conditions},\ }\href
  {https://doi.org/10.1103/PhysRevB.106.L180101} {\bibfield  {journal}
  {\bibinfo  {journal} {Phys. Rev. B}\ }\textbf {\bibinfo {volume} {106}},\
  \bibinfo {pages} {L180101} (\bibinfo {year} {2022})}\BibitemShut {NoStop}%
\bibitem [{\citenamefont {Morris}\ \emph {et~al.}(1994)\citenamefont {Morris},
  \citenamefont {Wang}, \citenamefont {Ho},\ and\ \citenamefont
  {Chan}}]{tpcMorris1994}%
  \BibitemOpen
  \bibfield  {author} {\bibinfo {author} {\bibfnamefont {J.~R.}\ \bibnamefont
  {Morris}}, \bibinfo {author} {\bibfnamefont {C.~Z.}\ \bibnamefont {Wang}},
  \bibinfo {author} {\bibfnamefont {K.~M.}\ \bibnamefont {Ho}},\ and\ \bibinfo
  {author} {\bibfnamefont {C.~T.}\ \bibnamefont {Chan}},\ }\bibfield  {title}
  {\bibinfo {title} {Melting line of aluminum from simulations of coexisting
  phases},\ }\href {https://doi.org/10.1103/PhysRevB.49.3109} {\bibfield
  {journal} {\bibinfo  {journal} {Phys. Rev. B}\ }\textbf {\bibinfo {volume}
  {49}},\ \bibinfo {pages} {3109} (\bibinfo {year} {1994})}\BibitemShut
  {NoStop}%
\bibitem [{\citenamefont {Lechner}\ and\ \citenamefont
  {Dellago}(2008)}]{pfaLechner2008}%
  \BibitemOpen
  \bibfield  {author} {\bibinfo {author} {\bibfnamefont {W.}~\bibnamefont
  {Lechner}}\ and\ \bibinfo {author} {\bibfnamefont {C.}~\bibnamefont
  {Dellago}},\ }\bibfield  {title} {\bibinfo {title} {Accurate determination of
  crystal structures based on averaged local bond order parameters},\ }\href
  {https://doi.org/10.1063/1.2977970} {\bibfield  {journal} {\bibinfo
  {journal} {J. Chem. Phys.}\ }\textbf {\bibinfo {volume} {129}},\ \bibinfo
  {pages} {114707} (\bibinfo {year} {2008})}\BibitemShut {NoStop}%
\bibitem [{\citenamefont {Li}\ \emph {et~al.}(2020)\citenamefont {Li},
  \citenamefont {Wu}, \citenamefont {Li}, \citenamefont {Xue}, \citenamefont
  {Tan}, \citenamefont {Zhou}, \citenamefont {Zhang}, \citenamefont {Xiong},
  \citenamefont {Gao},\ and\ \citenamefont {Sekine}}]{emcLi2020}%
  \BibitemOpen
  \bibfield  {author} {\bibinfo {author} {\bibfnamefont {J.}~\bibnamefont
  {Li}}, \bibinfo {author} {\bibfnamefont {Q.}~\bibnamefont {Wu}}, \bibinfo
  {author} {\bibfnamefont {J.}~\bibnamefont {Li}}, \bibinfo {author}
  {\bibfnamefont {T.}~\bibnamefont {Xue}}, \bibinfo {author} {\bibfnamefont
  {Y.}~\bibnamefont {Tan}}, \bibinfo {author} {\bibfnamefont {X.}~\bibnamefont
  {Zhou}}, \bibinfo {author} {\bibfnamefont {Y.}~\bibnamefont {Zhang}},
  \bibinfo {author} {\bibfnamefont {Z.}~\bibnamefont {Xiong}}, \bibinfo
  {author} {\bibfnamefont {Z.}~\bibnamefont {Gao}},\ and\ \bibinfo {author}
  {\bibfnamefont {T.}~\bibnamefont {Sekine}},\ }\bibfield  {title} {\bibinfo
  {title} {Shock melting curve of iron: A consensus on the temperature at the
  earth's inner core boundary},\ }\href
  {https://doi.org/https://doi.org/10.1029/2020GL087758} {\bibfield  {journal}
  {\bibinfo  {journal} {Geophys. Res. Lett.}\ }\textbf {\bibinfo {volume}
  {47}},\ \bibinfo {pages} {e2020GL087758} (\bibinfo {year}
  {2020})}\BibitemShut {NoStop}%
\bibitem [{sim(1929)}]{simonglatz}%
  \BibitemOpen
  \bibfield  {title} {\bibinfo {title} {Bemerkungen zur schmelzdrukkurve},\
  }\href@noop {} {\bibfield  {journal} {\bibinfo  {journal} {Z. Anorg. Allg.
  Chem.}\ }\textbf {\bibinfo {volume} {178}},\ \bibinfo {pages} {309} (\bibinfo
  {year} {1929})}\BibitemShut {NoStop}%
\bibitem [{\citenamefont {Jung}\ \emph {et~al.}(2023)\citenamefont {Jung},
  \citenamefont {Srinivasan}, \citenamefont {Forslund},\ and\ \citenamefont
  {Grabowski}}]{fepJung2023}%
  \BibitemOpen
  \bibfield  {author} {\bibinfo {author} {\bibfnamefont {J.~H.}\ \bibnamefont
  {Jung}}, \bibinfo {author} {\bibfnamefont {P.}~\bibnamefont {Srinivasan}},
  \bibinfo {author} {\bibfnamefont {A.}~\bibnamefont {Forslund}},\ and\
  \bibinfo {author} {\bibfnamefont {B.}~\bibnamefont {Grabowski}},\ }\bibfield
  {title} {\bibinfo {title} {High-accuracy thermodynamic properties to the
  melting point from ab initio calculations aided by machine-learning
  potentials},\ }\href {https://doi.org/10.1038/s41524-022-00956-8} {\bibfield
  {journal} {\bibinfo  {journal} {npj Comput. Mater.}\ }\textbf {\bibinfo
  {volume} {9}},\ \bibinfo {pages} {3} (\bibinfo {year} {2023})}\BibitemShut
  {NoStop}%
\bibitem [{\citenamefont {Gavriliuk}\ \emph {et~al.}(2023)\citenamefont
  {Gavriliuk}, \citenamefont {Struzhkin}, \citenamefont {Ivanova},
  \citenamefont {Prakapenka}, \citenamefont {Mironovich}, \citenamefont
  {Aksenov}, \citenamefont {Troyan},\ and\ \citenamefont
  {Morgenroth}}]{chsGavriliuk2023}%
  \BibitemOpen
  \bibfield  {author} {\bibinfo {author} {\bibfnamefont {A.~G.}\ \bibnamefont
  {Gavriliuk}}, \bibinfo {author} {\bibfnamefont {V.~V.}\ \bibnamefont
  {Struzhkin}}, \bibinfo {author} {\bibfnamefont {A.~G.}\ \bibnamefont
  {Ivanova}}, \bibinfo {author} {\bibfnamefont {V.~B.}\ \bibnamefont
  {Prakapenka}}, \bibinfo {author} {\bibfnamefont {A.~A.}\ \bibnamefont
  {Mironovich}}, \bibinfo {author} {\bibfnamefont {S.~N.}\ \bibnamefont
  {Aksenov}}, \bibinfo {author} {\bibfnamefont {I.~A.}\ \bibnamefont
  {Troyan}},\ and\ \bibinfo {author} {\bibfnamefont {W.}~\bibnamefont
  {Morgenroth}},\ }\bibfield  {title} {\bibinfo {title} {The first-order
  structural transition in \text{NiO} at high pressure},\ }\href
  {https://doi.org/10.1038/s42005-022-01098-5} {\bibfield  {journal} {\bibinfo
  {journal} {Commun. Phys.}\ }\textbf {\bibinfo {volume} {6}},\ \bibinfo
  {pages} {23} (\bibinfo {year} {2023})}\BibitemShut {NoStop}%
\bibitem [{\citenamefont {Ho}\ \emph {et~al.}(2024)\citenamefont {Ho},
  \citenamefont {Zhang}, \citenamefont {Haule}, \citenamefont {Jackson},
  \citenamefont {Dobrosavljevi{\'{c}}},\ and\ \citenamefont
  {Dobrosavljevic}}]{chsHo2024}%
  \BibitemOpen
  \bibfield  {author} {\bibinfo {author} {\bibfnamefont {W.-G.~D.}\
  \bibnamefont {Ho}}, \bibinfo {author} {\bibfnamefont {P.}~\bibnamefont
  {Zhang}}, \bibinfo {author} {\bibfnamefont {K.}~\bibnamefont {Haule}},
  \bibinfo {author} {\bibfnamefont {J.~M.}\ \bibnamefont {Jackson}}, \bibinfo
  {author} {\bibfnamefont {V.}~\bibnamefont {Dobrosavljevi{\'{c}}}},\ and\
  \bibinfo {author} {\bibfnamefont {V.~V.}\ \bibnamefont {Dobrosavljevic}},\
  }\bibfield  {title} {\bibinfo {title} {Quantum critical phase of \text{FeO}
  spans conditions of earth's lower mantle},\ }\href
  {https://doi.org/10.1038/s41467-024-47489-w} {\bibfield  {journal} {\bibinfo
  {journal} {Nat. Commun.}\ }\textbf {\bibinfo {volume} {15}},\ \bibinfo
  {pages} {3461} (\bibinfo {year} {2024})}\BibitemShut {NoStop}%
\bibitem [{\citenamefont {Ruban}\ \emph {et~al.}(2013)\citenamefont {Ruban},
  \citenamefont {Belonoshko},\ and\ \citenamefont
  {Skorodumova}}]{PhysRevB.87.014405}%
  \BibitemOpen
  \bibfield  {author} {\bibinfo {author} {\bibfnamefont {A.~V.}\ \bibnamefont
  {Ruban}}, \bibinfo {author} {\bibfnamefont {A.~B.}\ \bibnamefont
  {Belonoshko}},\ and\ \bibinfo {author} {\bibfnamefont {N.~V.}\ \bibnamefont
  {Skorodumova}},\ }\bibfield  {title} {\bibinfo {title} {Impact of magnetism
  on \text{Fe} under earth's core conditions},\ }\href
  {https://doi.org/10.1103/PhysRevB.87.014405} {\bibfield  {journal} {\bibinfo
  {journal} {Phys. Rev. B}\ }\textbf {\bibinfo {volume} {87}},\ \bibinfo
  {pages} {014405} (\bibinfo {year} {2013})}\BibitemShut {NoStop}%
\bibitem [{\citenamefont {Konopkova}\ \emph {et~al.}(2026)\citenamefont
  {Konopkova} \emph {et~al.}}]{emcKonopkova2025}%
  \BibitemOpen
  \bibfield  {author} {\bibinfo {author} {\bibfnamefont {Z.}~\bibnamefont
  {Konopkova}} \emph {et~al.},\ }\bibfield  {title} {\bibinfo {title}
  {Observation of body-centered cubic iron above 200 gigapascals},\ }\href
  {https://journals.aps.org/prb/accepted/10.1103/kxnf-6c5y} {\bibfield
  {journal} {\bibinfo  {journal} {Phys. Rev. B}\ } (\bibinfo {year} {2026})},\
  \bibinfo {note} {in press; accepted 15 July 2026},\ \Eprint
  {https://arxiv.org/abs/2505.15397} {arXiv:2505.15397} \BibitemShut {NoStop}%
\bibitem [{\citenamefont {Blaha}\ \emph {et~al.}(2020)\citenamefont {Blaha},
  \citenamefont {Schwarz}, \citenamefont {Tran}, \citenamefont {Laskowski},
  \citenamefont {Madsen},\ and\ \citenamefont {Marks}}]{wien2kblaha2020}%
  \BibitemOpen
  \bibfield  {author} {\bibinfo {author} {\bibfnamefont {P.}~\bibnamefont
  {Blaha}}, \bibinfo {author} {\bibfnamefont {K.}~\bibnamefont {Schwarz}},
  \bibinfo {author} {\bibfnamefont {F.}~\bibnamefont {Tran}}, \bibinfo {author}
  {\bibfnamefont {R.}~\bibnamefont {Laskowski}}, \bibinfo {author}
  {\bibfnamefont {G.~K.~H.}\ \bibnamefont {Madsen}},\ and\ \bibinfo {author}
  {\bibfnamefont {L.~D.}\ \bibnamefont {Marks}},\ }\bibfield  {title} {\bibinfo
  {title} {{WIEN2k: An APW+lo program for calculating the properties of
  solids}},\ }\href {https://doi.org/10.1063/1.5143061} {\bibfield  {journal}
  {\bibinfo  {journal} {J. Chem. Phys.}\ }\textbf {\bibinfo {volume} {152}},\
  \bibinfo {pages} {074101} (\bibinfo {year} {2020})}\BibitemShut {NoStop}%
\bibitem [{\citenamefont {Haule}\ \emph {et~al.}(2010)\citenamefont {Haule},
  \citenamefont {Yee},\ and\ \citenamefont {Kim}}]{edmftfHaule2010}%
  \BibitemOpen
  \bibfield  {author} {\bibinfo {author} {\bibfnamefont {K.}~\bibnamefont
  {Haule}}, \bibinfo {author} {\bibfnamefont {C.-H.}\ \bibnamefont {Yee}},\
  and\ \bibinfo {author} {\bibfnamefont {K.}~\bibnamefont {Kim}},\ }\bibfield
  {title} {\bibinfo {title} {Dynamical mean-field theory within the
  full-potential methods: Electronic structure of ${\text{ceirin}}_{5}$,
  ${\text{cecoin}}_{5}$, and ${\text{cerhin}}_{5}$},\ }\href
  {https://doi.org/10.1103/PhysRevB.81.195107} {\bibfield  {journal} {\bibinfo
  {journal} {Phys. Rev. B}\ }\textbf {\bibinfo {volume} {81}},\ \bibinfo
  {pages} {195107} (\bibinfo {year} {2010})}\BibitemShut {NoStop}%
\bibitem [{\citenamefont {Haule}\ and\ \citenamefont
  {Birol}(2015)}]{edmftfHaule2015}%
  \BibitemOpen
  \bibfield  {author} {\bibinfo {author} {\bibfnamefont {K.}~\bibnamefont
  {Haule}}\ and\ \bibinfo {author} {\bibfnamefont {T.}~\bibnamefont {Birol}},\
  }\bibfield  {title} {\bibinfo {title} {Free energy from stationary
  implementation of the $\mathrm{DFT}+\mathrm{DMFT}$ functional},\ }\href
  {https://doi.org/10.1103/PhysRevLett.115.256402} {\bibfield  {journal}
  {\bibinfo  {journal} {Phys. Rev. Lett.}\ }\textbf {\bibinfo {volume} {115}},\
  \bibinfo {pages} {256402} (\bibinfo {year} {2015})}\BibitemShut {NoStop}%
\bibitem [{\citenamefont {Haule}(2007)}]{edmftfHaule2007}%
  \BibitemOpen
  \bibfield  {author} {\bibinfo {author} {\bibfnamefont {K.}~\bibnamefont
  {Haule}},\ }\bibfield  {title} {\bibinfo {title} {Quantum monte carlo
  impurity solver for cluster dynamical mean-field theory and electronic
  structure calculations with adjustable cluster base},\ }\href
  {https://doi.org/10.1103/PhysRevB.75.155113} {\bibfield  {journal} {\bibinfo
  {journal} {Phys. Rev. B}\ }\textbf {\bibinfo {volume} {75}},\ \bibinfo
  {pages} {155113} (\bibinfo {year} {2007})}\BibitemShut {NoStop}%
\bibitem [{\citenamefont {Haule}\ and\ \citenamefont
  {Pascut}(2016)}]{edmftfHaule2016}%
  \BibitemOpen
  \bibfield  {author} {\bibinfo {author} {\bibfnamefont {K.}~\bibnamefont
  {Haule}}\ and\ \bibinfo {author} {\bibfnamefont {G.~L.}\ \bibnamefont
  {Pascut}},\ }\bibfield  {title} {\bibinfo {title} {Forces for structural
  optimizations in correlated materials within a dft+embedded dmft functional
  approach},\ }\href {https://doi.org/10.1103/PhysRevB.94.195146} {\bibfield
  {journal} {\bibinfo  {journal} {Phys. Rev. B}\ }\textbf {\bibinfo {volume}
  {94}},\ \bibinfo {pages} {195146} (\bibinfo {year} {2016})}\BibitemShut
  {NoStop}%
\bibitem [{\citenamefont {Haule}\ \emph {et~al.}(2014)\citenamefont {Haule},
  \citenamefont {Birol},\ and\ \citenamefont {Kotliar}}]{dcHaule2014}%
  \BibitemOpen
  \bibfield  {author} {\bibinfo {author} {\bibfnamefont {K.}~\bibnamefont
  {Haule}}, \bibinfo {author} {\bibfnamefont {T.}~\bibnamefont {Birol}},\ and\
  \bibinfo {author} {\bibfnamefont {G.}~\bibnamefont {Kotliar}},\ }\bibfield
  {title} {\bibinfo {title} {Covalency in transition-metal oxides within
  all-electron dynamical mean-field theory},\ }\href
  {https://doi.org/10.1103/PhysRevB.90.075136} {\bibfield  {journal} {\bibinfo
  {journal} {Phys. Rev. B}\ }\textbf {\bibinfo {volume} {90}},\ \bibinfo
  {pages} {075136} (\bibinfo {year} {2014})}\BibitemShut {NoStop}%
\bibitem [{\citenamefont {Kresse}\ and\ \citenamefont
  {Furthmüller}(1996)}]{vaspKresse1996}%
  \BibitemOpen
  \bibfield  {author} {\bibinfo {author} {\bibfnamefont {G.}~\bibnamefont
  {Kresse}}\ and\ \bibinfo {author} {\bibfnamefont {J.}~\bibnamefont
  {Furthmüller}},\ }\bibfield  {title} {\bibinfo {title} {Efficiency of
  ab-initio total energy calculations for metals and semiconductors using a
  plane-wave basis set},\ }\href
  {https://doi.org/https://doi.org/10.1016/0927-0256(96)00008-0} {\bibfield
  {journal} {\bibinfo  {journal} {Computational Materials Science}\ }\textbf
  {\bibinfo {volume} {6}},\ \bibinfo {pages} {15} (\bibinfo {year}
  {1996})}\BibitemShut {NoStop}%
\bibitem [{\citenamefont {Kresse}\ and\ \citenamefont
  {Joubert}(1999)}]{vaspKresse1999}%
  \BibitemOpen
  \bibfield  {author} {\bibinfo {author} {\bibfnamefont {G.}~\bibnamefont
  {Kresse}}\ and\ \bibinfo {author} {\bibfnamefont {D.}~\bibnamefont
  {Joubert}},\ }\bibfield  {title} {\bibinfo {title} {From ultrasoft
  pseudopotentials to the projector augmented-wave method},\ }\href
  {https://doi.org/10.1103/PhysRevB.59.1758} {\bibfield  {journal} {\bibinfo
  {journal} {Phys. Rev. B}\ }\textbf {\bibinfo {volume} {59}},\ \bibinfo
  {pages} {1758} (\bibinfo {year} {1999})}\BibitemShut {NoStop}%
\bibitem [{\citenamefont {Chen}\ and\ \citenamefont {Ong}(2022)}]{m3gChen2022}%
  \BibitemOpen
  \bibfield  {author} {\bibinfo {author} {\bibfnamefont {C.}~\bibnamefont
  {Chen}}\ and\ \bibinfo {author} {\bibfnamefont {S.~P.}\ \bibnamefont {Ong}},\
  }\bibfield  {title} {\bibinfo {title} {A universal graph deep learning
  interatomic potential for the periodic table},\ }\href
  {https://doi.org/10.1038/s43588-022-00349-3} {\bibfield  {journal} {\bibinfo
  {journal} {Nat. Comp. Sci.}\ }\textbf {\bibinfo {volume} {2}},\ \bibinfo
  {pages} {718} (\bibinfo {year} {2022})}\BibitemShut {NoStop}%
\bibitem [{\citenamefont {Ozawa}\ \emph {et~al.}(2011)\citenamefont {Ozawa},
  \citenamefont {Takahashi}, \citenamefont {Hirose}, \citenamefont {Ohishi},\
  and\ \citenamefont {Hirao}}]{feoOzawa2011}%
  \BibitemOpen
  \bibfield  {author} {\bibinfo {author} {\bibfnamefont {H.}~\bibnamefont
  {Ozawa}}, \bibinfo {author} {\bibfnamefont {F.}~\bibnamefont {Takahashi}},
  \bibinfo {author} {\bibfnamefont {K.}~\bibnamefont {Hirose}}, \bibinfo
  {author} {\bibfnamefont {Y.}~\bibnamefont {Ohishi}},\ and\ \bibinfo {author}
  {\bibfnamefont {N.}~\bibnamefont {Hirao}},\ }\bibfield  {title} {\bibinfo
  {title} {Phase transition of \text{FeO} and stratification in earth’s outer
  core},\ }\href {https://doi.org/10.1126/science.1208265} {\bibfield
  {journal} {\bibinfo  {journal} {Science}\ }\textbf {\bibinfo {volume}
  {334}},\ \bibinfo {pages} {792} (\bibinfo {year} {2011})}\BibitemShut
  {NoStop}%
\bibitem [{\citenamefont {Rao}\ and\ \citenamefont {Zhu}(2024)}]{Rao2024}%
  \BibitemOpen
  \bibfield  {author} {\bibinfo {author} {\bibfnamefont {R.}~\bibnamefont
  {Rao}}\ and\ \bibinfo {author} {\bibfnamefont {L.}~\bibnamefont {Zhu}},\
  }\bibfield  {title} {\bibinfo {title} {Phase transitions of correlated
  systems from graph neural networks with quantum embedding techniques},\
  }\href {https://doi.org/10.1103/PhysRevB.110.245111} {\bibfield  {journal}
  {\bibinfo  {journal} {Phys. Rev. B}\ }\textbf {\bibinfo {volume} {110}},\
  \bibinfo {pages} {245111} (\bibinfo {year} {2024})}\BibitemShut {NoStop}%
\bibitem [{\citenamefont {Parcollet}\ \emph {et~al.}(2015)\citenamefont
  {Parcollet}, \citenamefont {Ferrero}, \citenamefont {Ayral}, \citenamefont
  {Hafermann}, \citenamefont {Krivenko}, \citenamefont {Messio},\ and\
  \citenamefont {Seth}}]{Parcollet2015}%
  \BibitemOpen
  \bibfield  {author} {\bibinfo {author} {\bibfnamefont {O.}~\bibnamefont
  {Parcollet}}, \bibinfo {author} {\bibfnamefont {M.}~\bibnamefont {Ferrero}},
  \bibinfo {author} {\bibfnamefont {T.}~\bibnamefont {Ayral}}, \bibinfo
  {author} {\bibfnamefont {H.}~\bibnamefont {Hafermann}}, \bibinfo {author}
  {\bibfnamefont {I.}~\bibnamefont {Krivenko}}, \bibinfo {author}
  {\bibfnamefont {L.}~\bibnamefont {Messio}},\ and\ \bibinfo {author}
  {\bibfnamefont {P.}~\bibnamefont {Seth}},\ }\bibfield  {title} {\bibinfo
  {title} {{TRIQS: A toolbox for research on interacting quantum systems}},\
  }\href {https://doi.org/10.1016/j.cpc.2015.04.023} {\bibfield  {journal}
  {\bibinfo  {journal} {Comput. Phys. Commun.}\ }\textbf {\bibinfo {volume}
  {196}},\ \bibinfo {pages} {398} (\bibinfo {year} {2015})}\BibitemShut
  {NoStop}%
\bibitem [{\citenamefont {Loshchilov}\ and\ \citenamefont
  {Hutter}()}]{adamwLoshchilov2019}%
  \BibitemOpen
  \bibfield  {author} {\bibinfo {author} {\bibfnamefont {I.}~\bibnamefont
  {Loshchilov}}\ and\ \bibinfo {author} {\bibfnamefont {F.}~\bibnamefont
  {Hutter}},\ }\href {https://arxiv.org/abs/1711.05101} {\bibinfo {title}
  {Decoupled weight decay regularization}},\ \Eprint
  {https://arxiv.org/abs/1711.05101} {arXiv:1711.05101} \BibitemShut {NoStop}%
\bibitem [{\citenamefont {Ziegler}\ and\ \citenamefont
  {Biersack}(1985)}]{zblZiegler1985}%
  \BibitemOpen
  \bibfield  {author} {\bibinfo {author} {\bibfnamefont {J.~F.}\ \bibnamefont
  {Ziegler}}\ and\ \bibinfo {author} {\bibfnamefont {J.~P.}\ \bibnamefont
  {Biersack}},\ }\bibinfo {title} {The stopping and range of ions in matter},\
  in\ \href {https://doi.org/10.1007/978-1-4615-8103-1_3} {\emph {\bibinfo
  {booktitle} {Treatise on Heavy-Ion Science: Volume 6: Astrophysics,
  Chemistry, and Condensed Matter}}},\ \bibinfo {editor} {edited by\ \bibinfo
  {editor} {\bibfnamefont {D.~A.}\ \bibnamefont {Bromley}}}\ (\bibinfo
  {publisher} {Springer US},\ \bibinfo {address} {Boston, MA},\ \bibinfo {year}
  {1985})\ pp.\ \bibinfo {pages} {93--129}\BibitemShut {NoStop}%
\bibitem [{\citenamefont {Larsen}\ \emph {et~al.}(2017)\citenamefont {Larsen},
  \citenamefont {Mortensen}, \citenamefont {Blomqvist}, \citenamefont
  {Castelli}, \citenamefont {Christensen}, \citenamefont {Dułak},
  \citenamefont {Friis}, \citenamefont {Groves}, \citenamefont {Hammer},
  \citenamefont {Hargus}, \citenamefont {Hermes}, \citenamefont {Jennings},
  \citenamefont {Jensen}, \citenamefont {Kermode}, \citenamefont {Kitchin},
  \citenamefont {Kolsbjerg}, \citenamefont {Kubal}, \citenamefont {Kaasbjerg},
  \citenamefont {Lysgaard}, \citenamefont {Maronsson}, \citenamefont {Maxson},
  \citenamefont {Olsen}, \citenamefont {Pastewka}, \citenamefont {Peterson},
  \citenamefont {Rostgaard}, \citenamefont {Schiøtz}, \citenamefont {Schütt},
  \citenamefont {Strange}, \citenamefont {Thygesen}, \citenamefont {Vegge},
  \citenamefont {Vilhelmsen}, \citenamefont {Walter}, \citenamefont {Zeng},\
  and\ \citenamefont {Jacobsen}}]{aseLarsen2017}%
  \BibitemOpen
  \bibfield  {author} {\bibinfo {author} {\bibfnamefont {A.~H.}\ \bibnamefont
  {Larsen}}, \bibinfo {author} {\bibfnamefont {J.~J.}\ \bibnamefont
  {Mortensen}}, \bibinfo {author} {\bibfnamefont {J.}~\bibnamefont
  {Blomqvist}}, \bibinfo {author} {\bibfnamefont {I.~E.}\ \bibnamefont
  {Castelli}}, \bibinfo {author} {\bibfnamefont {R.}~\bibnamefont
  {Christensen}}, \bibinfo {author} {\bibfnamefont {M.}~\bibnamefont {Dułak}},
  \bibinfo {author} {\bibfnamefont {J.}~\bibnamefont {Friis}}, \bibinfo
  {author} {\bibfnamefont {M.~N.}\ \bibnamefont {Groves}}, \bibinfo {author}
  {\bibfnamefont {B.}~\bibnamefont {Hammer}}, \bibinfo {author} {\bibfnamefont
  {C.}~\bibnamefont {Hargus}}, \bibinfo {author} {\bibfnamefont {E.~D.}\
  \bibnamefont {Hermes}}, \bibinfo {author} {\bibfnamefont {P.~C.}\
  \bibnamefont {Jennings}}, \bibinfo {author} {\bibfnamefont {P.~B.}\
  \bibnamefont {Jensen}}, \bibinfo {author} {\bibfnamefont {J.}~\bibnamefont
  {Kermode}}, \bibinfo {author} {\bibfnamefont {J.~R.}\ \bibnamefont
  {Kitchin}}, \bibinfo {author} {\bibfnamefont {E.~L.}\ \bibnamefont
  {Kolsbjerg}}, \bibinfo {author} {\bibfnamefont {J.}~\bibnamefont {Kubal}},
  \bibinfo {author} {\bibfnamefont {K.}~\bibnamefont {Kaasbjerg}}, \bibinfo
  {author} {\bibfnamefont {S.}~\bibnamefont {Lysgaard}}, \bibinfo {author}
  {\bibfnamefont {J.~B.}\ \bibnamefont {Maronsson}}, \bibinfo {author}
  {\bibfnamefont {T.}~\bibnamefont {Maxson}}, \bibinfo {author} {\bibfnamefont
  {T.}~\bibnamefont {Olsen}}, \bibinfo {author} {\bibfnamefont
  {L.}~\bibnamefont {Pastewka}}, \bibinfo {author} {\bibfnamefont
  {A.}~\bibnamefont {Peterson}}, \bibinfo {author} {\bibfnamefont
  {C.}~\bibnamefont {Rostgaard}}, \bibinfo {author} {\bibfnamefont
  {J.}~\bibnamefont {Schiøtz}}, \bibinfo {author} {\bibfnamefont
  {O.}~\bibnamefont {Schütt}}, \bibinfo {author} {\bibfnamefont
  {M.}~\bibnamefont {Strange}}, \bibinfo {author} {\bibfnamefont {K.~S.}\
  \bibnamefont {Thygesen}}, \bibinfo {author} {\bibfnamefont {T.}~\bibnamefont
  {Vegge}}, \bibinfo {author} {\bibfnamefont {L.}~\bibnamefont {Vilhelmsen}},
  \bibinfo {author} {\bibfnamefont {M.}~\bibnamefont {Walter}}, \bibinfo
  {author} {\bibfnamefont {Z.}~\bibnamefont {Zeng}},\ and\ \bibinfo {author}
  {\bibfnamefont {K.~W.}\ \bibnamefont {Jacobsen}},\ }\bibfield  {title}
  {\bibinfo {title} {The atomic simulation environment—a python library for
  working with atoms},\ }\href
  {http://stacks.iop.org/0953-8984/29/i=27/a=273002} {\bibfield  {journal}
  {\bibinfo  {journal} {J. Phys. Condens. Matter}\ }\textbf {\bibinfo {volume}
  {29}},\ \bibinfo {pages} {273002} (\bibinfo {year} {2017})}\BibitemShut
  {NoStop}%
\bibitem [{\citenamefont {Hirao}\ \emph {et~al.}(2022)\citenamefont {Hirao},
  \citenamefont {Akahama},\ and\ \citenamefont {Ohishi}}]{felpHirao2022}%
  \BibitemOpen
  \bibfield  {author} {\bibinfo {author} {\bibfnamefont {N.}~\bibnamefont
  {Hirao}}, \bibinfo {author} {\bibfnamefont {Y.}~\bibnamefont {Akahama}},\
  and\ \bibinfo {author} {\bibfnamefont {Y.}~\bibnamefont {Ohishi}},\
  }\bibfield  {title} {\bibinfo {title} {Equations of state of iron and nickel
  to the pressure at the center of the earth},\ }\href
  {https://doi.org/10.1063/5.0074340} {\bibfield  {journal} {\bibinfo
  {journal} {Matter Radiat. Extremes}\ }\textbf {\bibinfo {volume} {7}},\
  \bibinfo {pages} {038403} (\bibinfo {year} {2022})}\BibitemShut {NoStop}%
\end{thebibliography}%

\end{document}